\documentclass[aip,a4paper,reprint,twoside]{revtex4-1}

\usepackage{filecontents}

\usepackage[english]{babel}
\usepackage[utf8]{inputenc}
\usepackage[labelfont=up]{subcaption}
\usepackage[x11names]{xcolor}
\usepackage{float}
\usepackage[colorinlistoftodos, color=green!40, prependcaption]{todonotes}
\usepackage[pdftex, pdftitle={Article}, pdfauthor={Author}]{hyperref}
\hypersetup{colorlinks=true, citecolor=black, linkcolor=black, urlcolor=blue} 
\usepackage{mathtools}
\usepackage{etoolbox}
\usepackage[top=1.in, bottom=1.in, left=0.5in, right=0.5in]{geometry}
\usepackage{amsthm, amssymb}
\usepackage{graphicx}
	\graphicspath{{img/}}
\usepackage{lipsum}

\DeclarePairedDelimiter\ton{(}{)}
\DeclarePairedDelimiter\qua{[}{]}
\DeclarePairedDelimiter\gra{\{}{\}}
\DeclarePairedDelimiter\mean{\langle}{\rangle}

\newcommand{\me}{\mathrm{e}}

\newcommand{\be}{\begin{equation}}
\newcommand{\ee}{\end{equation}}

\makeatletter
\def\@email#1#2{%
 \endgroup
 \patchcmd{\titleblock@produce}
  {\frontmatter@RRAPformat}
  {\frontmatter@RRAPformat{\produce@RRAP{*#1\href{mailto:#2}{#2}}}\frontmatter@RRAPformat}
  {}{}
}%
\makeatother

\begin{document}

\title{Filter Bubble effect in the multistate voter model}

\author{Giulio Iannelli}
\affiliation{Centro Ricerche Enrico Fermi, Piazza del Viminale, 1, I-00184 Rome, Italy.}
\affiliation{Dipartimento di Fisica, Università di Roma “Tor Vergata”, 00133 Roma, Italy.}

\author{Giordano De Marzo$^*$}
\affiliation{Centro Ricerche Enrico Fermi, Piazza del Viminale, 1, I-00184 Rome, Italy.}
\affiliation{Dipartimento di Fisica Universit\`a ``Sapienza”, P.le A. Moro, 2, I-00185 Rome, Italy.}
\affiliation{Sapienza School for Advanced Studies, ``Sapienza'', P.le A. Moro, 2, I-00185 Rome, Italy.}

\author{Claudio Castellano}
\affiliation{Istituto dei Sistemi Complessi (ISC-CNR), Via dei Taurini, 19, I-00185 Rome, Italy}
\affiliation{Centro Ricerche Enrico Fermi, Piazza del Viminale, 1, I-00184 Rome, Italy.}
\email{giordano.demarzo@cref.it}

\date{\today} % Leave empty to omit a date
\begin{abstract}
  Social media influence online activity by recommending to users
  content strongly correlated with what they have preferred in the
  past. In this way they constrain users within filter bubbles that
  strongly limit their exposure to new or alternative content.  We
  investigate this type of dynamics by considering a multistate voter
  model where, with a given probability $\lambda$, a user interacts
  with a ``personalized information'' suggesting the opinion most
  frequently held in the past. By means of theoretical arguments and
  numerical simulations, we show the existence of a nontrivial
  transition between a region (for small $\lambda$) where consensus is
  reached and a region (above a threshold $\lambda_c$) where the
  system gets polarized and clusters of users with different opinions
  persist indefinitely. The threshold always vanishes for large system
  size $N$, showing that consensus becomes impossible for a large
  number of users. This finding opens new questions about the side
  effects of the widespread use of personalized recommendation algorithms.
\end{abstract}
%\keywords{first keyword, second keyword, third keyword}

\maketitle

\textbf{Information is nowadays mainly diffused via digital channels
  as people increasingly use online platforms instead of newspapers or
  television to access news. If, on the one hand, this makes information
  easily available to all, on the other hand it allows extremely detailed
  personalization.
  Most web sites use recommendation algorithms to provide users with content
  in line with their taste and way of thinking. Examples are the
  personalized page rank of Google, ``suggested for you'' posts by Facebook
  or the recommendations of Amazon and Netflix. This leads to the formation
  of the so-called ``filter bubbles'', in which users are exposed almost
  exclusively only to content they have already shown to be interested to.
  In this manuscript we model this phenomenon by a modification of a very
  simple model for opinion dynamics (the multi-state voter model)
  where individuals are influenced not only by their peers but also by
  an external ``field'’ which encodes information about the opinions
  the individual held in the past. This field, if large enough, leads
  to a polarized steady state in which users opinions are crystallized
  and consensus is no more possible.}

\section{Introduction}
The concept of ``filter bubble'' has crossed
the boundaries of the academic world reaching public discourse
and mainstream media. This reflects the realization that online social
media (OSM) have a tremendous impact on how people share information and
form their opinions. For this reason they may constitute not only a great
opportunity for the diffusion of knowledge, but also a great threat for
the stability of social fabric and the functioning of democracy.
A filter bubble occurs when a user is selectively exposed
predominantly to content that tends to reinforce his/her
current opinion/belief/state, while suppressing other alternatives \cite{pariser2011filter, dillahunt2015detecting, nagulendra2014understanding}.
In OSM this typically happens because of personalized
recommender systems, which leverage information on past user activity
to provide suggestions which, aiming at maximizing user satisfaction,
tend to be very similar to what the user has already shown to prefer \cite{nguyen2014exploring, bryant2020youtube, o2013extreme}.
Together with the ``echo chamber'' effect~\cite{Cinelli2021, Cota2019, barbera2015tweeting},
filter bubbles are thought to be at the heart of the overall increase of polarization
and radicalization that is observed in many social contexts~\cite{Pew2017, chitra2020analyzing, maes2015will}.

A great deal of activity has been devoted in the last years to
the goal of understanding what are the basic microscopic mechanisms
underlying the rise of polarization and how phenomena
observed at population scale are linked to
them~\cite{galam1997rational, crokidakis2013role, Ciampaglia2018,Perra2019,Sirbu2019, freitas2020imperfect, Baumann2020,Peralta2021,Baron2021}
These efforts follow the line of research aimed
at understanding how basic mechanisms underlying
the interaction of individuals give rise to collective consensus
phenomena~\cite{Castellano2009,Sen2014}.
The voter model~\cite{Clifford1973,Frachebourg1996,Rednerbook} played
an important role in this activity, because of its extremely simple
nature amenable to exact analytical treatment. Its dynamics describes
the evolution of a population of agents which have to choose between
two perfectly equivalent alternatives and do it by selecting at
random a peer and copying their selection.
Starting from a disordered state, clusters of individuals sharing the
same opinion form and grow over time. For any structure of the interaction
pattern among individuals, consensus (i.e. all agents having the same
opinion) is invariably reached.
A very natural generalization is the Multistate
Voter Model (MVM), where the number of available equivalent
options is a fixed value $M>2$~\cite{Starnini2012,Pickering2016,Peralta2020}.
In mean-field, MVM reaches consensus
for any value of $M$, and the average time required for it depends on $M$ only weakly.
In this paper we investigate the behavior of MVM in the presence
of an additional interaction mechanism which biases the opinion of an
agent toward the state the agent has chosen most frequently in the past.
This interaction mimics the filter bubble
effect of personalized recommenders in OSM, which present to users
suggestions based on their previous behavior.

The effect of personalized information on binary voter model dynamics
has been investigated in a previous publication~\cite{DeMarzo2020}.
In a homogeneous mean-field framework it was shown that for
sufficiently strong coupling with the agents' past, consensus is no
longer reached and the system remains stuck in a polarized state with
coexistence of both opinions.  Here we generalize the work in
Ref.~\cite{DeMarzo2020} by studying both analytically and numerically
the effect of personalized information in the context of the MVM.  The
goal is to understand whether, depending on the number of agents $N$,
the number of possible opinions $M$ and the strength of the
personalized information ($\lambda$, to be defined below) consensus is
reached or not.  Depending on the scaling of $M$ with respect to $N$
we identify three different regimes and in each of them we compute the
threshold $\lambda_c$ separating consensus from polarized
states. Numerical simulations are in good agreement with theory in all
cases expect for $M=N$ and they also show that the transition from
consensus to polarization is continuous and characterized by a power
law distribution of opinions numerosity at the threshold. Remarkably,
for $N\to\infty$ the threshold goes to zero in all three regimes and
this implies that in large systems even a very small form of
personalized information always breaks consensus, leading to opinion
polarization.

\section{Multistate voter model with personalized information}\label{secII}
\begin{figure*}
	\centering
	\begin{subfigure}{.45\textwidth}
		\includegraphics[width=\textwidth]{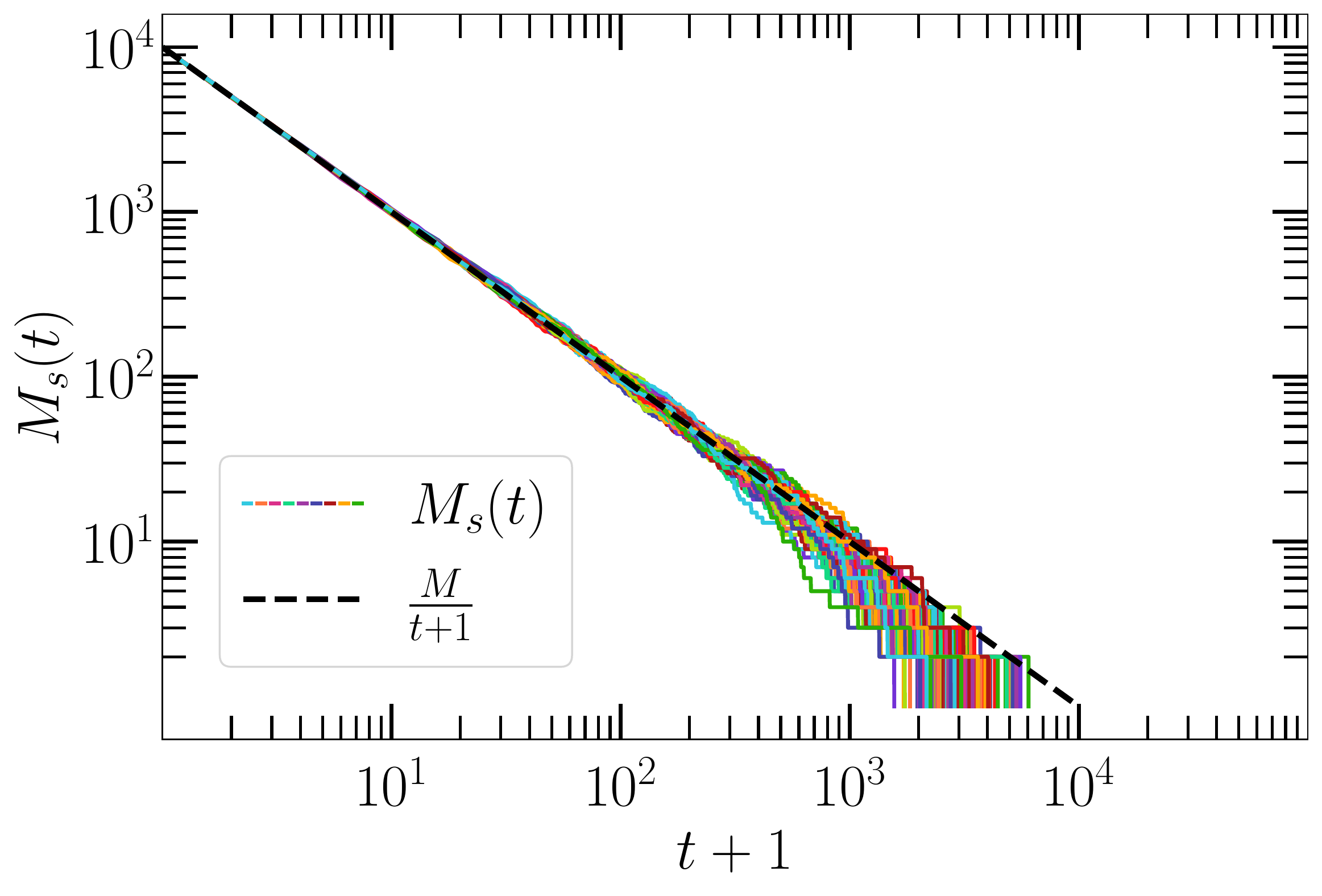}
		\caption{}\label{sfig:large_c_1}
	\end{subfigure}\hfill%
	\begin{subfigure}{.45\textwidth}
		\includegraphics[width=\textwidth]{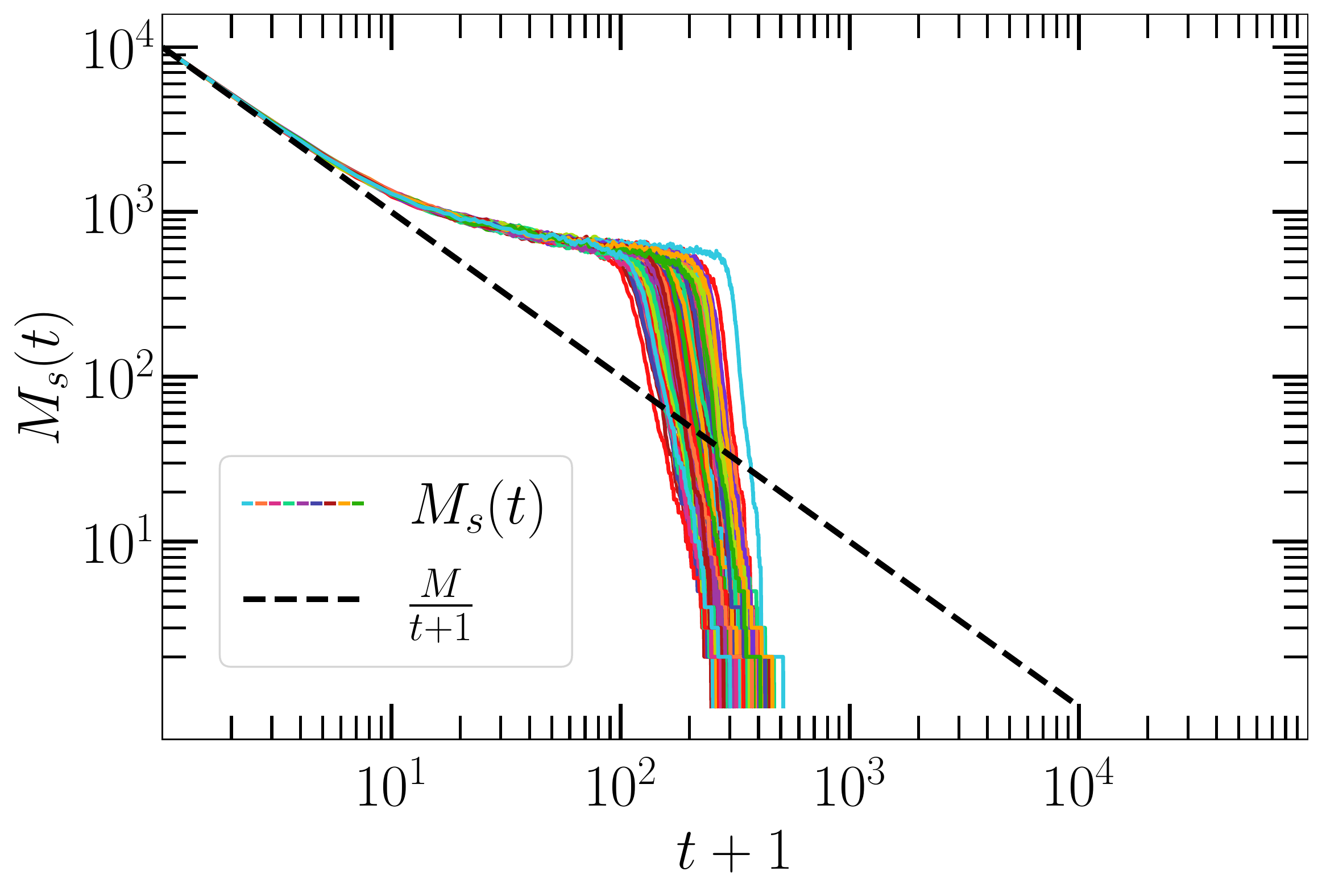}
		\caption{}\label{sfig:large_c_2}
	\end{subfigure}
	\begin{subfigure}{.45\textwidth}
		\includegraphics[width=\textwidth]{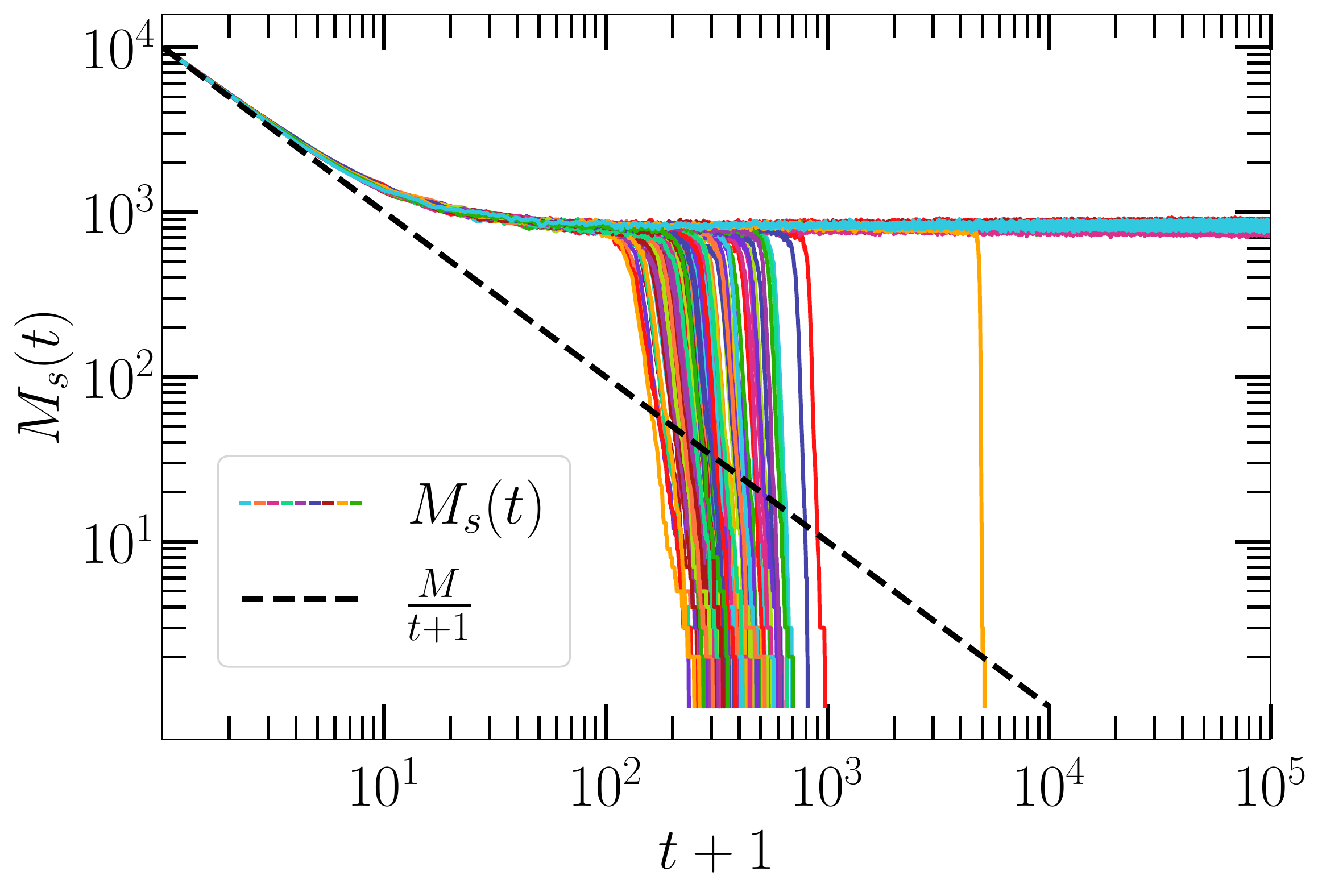}
		\caption{}\label{sfig:large_c_3}
	\end{subfigure}\hfill%
	\begin{subfigure}{.45\textwidth}
		\includegraphics[width=\textwidth]{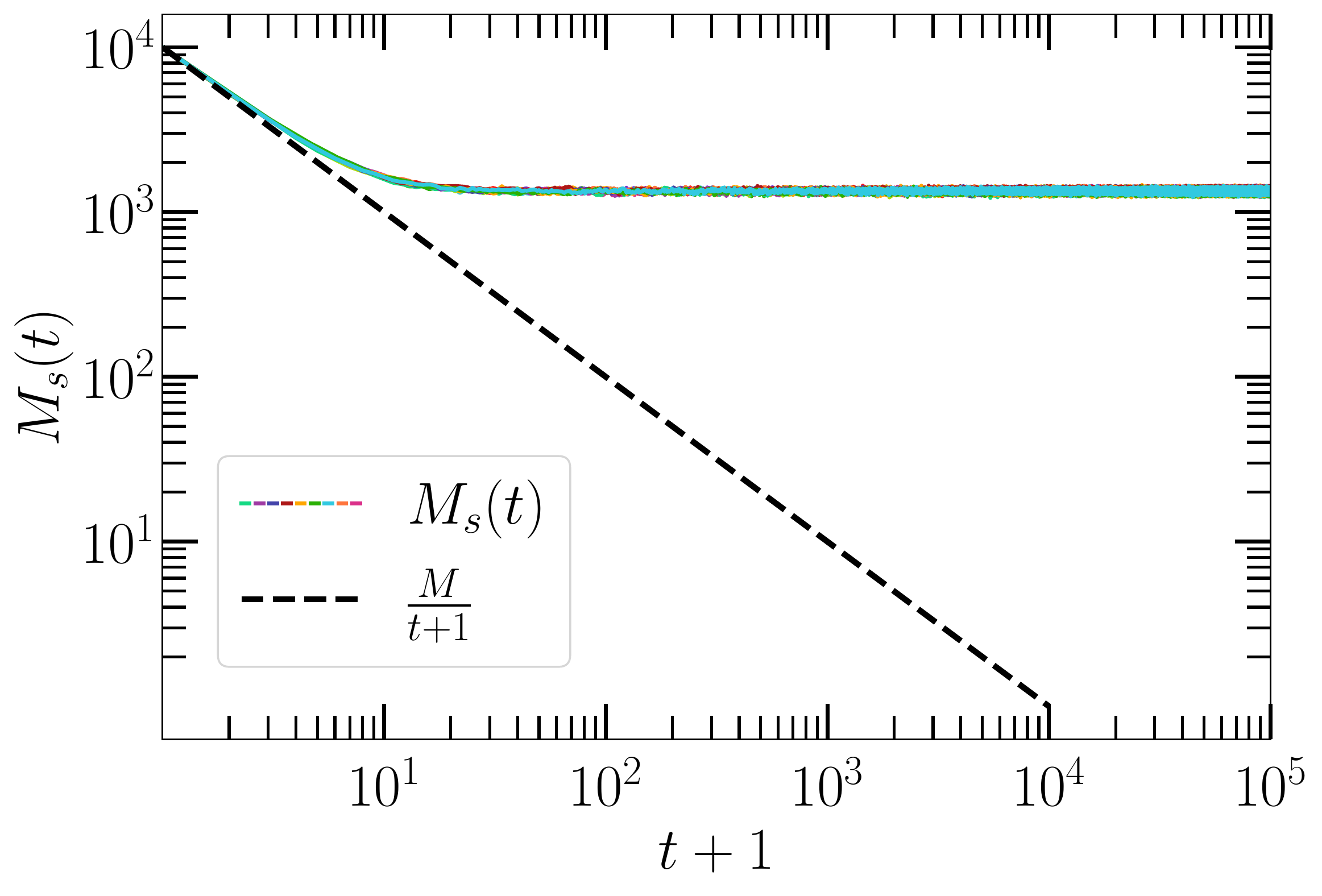}
		\caption{}\label{sfig:large_c_4}
	\end{subfigure}
	\caption{Evolution of the number of surviving opinions for 100
          realizations of the dynamical process with various values of
          $\lambda$. (a) Very small $\lambda=0.001$; (b) Slightly
          subcritical $\lambda=0.04$; (c) Critical $\lambda=0.05$; (d)
          Supercritical $\lambda=0.08$. In all panels $N=M=10000$,
          $c=\infty$.  For convenience we plot $t+1$ along the
          horizontal axis.}
 \label{fig:large_c}
\end{figure*}
\subsection{Definition of the model}
Many real life situations are characterized by the
presence of more than two possible opinions or
factions, as in the case of political elections or
football clubs, and so the voter model with binary
opinions is not suitable for schematizing the dynamics
of such systems. However this difficulty can be easily
bypassed by considering the usual voter dynamics and
allowing the states of the $N$ agents to vary among
$M$ distinct opinions instead of only two, so to
obtain the so called multistate voter model.
%This is in some sense equivalent to the Potts generalization
%of Ising spins. We will use the terms agents and spins interchangeably
%in the rest of the paper.
Denoting by $\sigma_i(t)$ the opinion of agent $i$ at time $t$ we have
$\sigma_i(t)=1\ldots M$ and the update rule reads
\begin{equation}
  \sigma_i(t+\delta t) = \sigma_j(t) \ \text{with prob.} \ \frac{A_{ij}}{\sum_j A_{ij}}.
  \label{eq:update_MVM}
\end{equation}
Here $A_{ij}$ is the binary adjacency matrix of the undirected network
over which the dynamics take place, while $\delta t=1/N$. In the
following we focus on the mean field case, meaning that the underlying
network is a complete graph and it holds $A_{ij}=1$ for $i\neq j$; in
this case the update rule Eq.~\eqref{eq:update_MVM} can be written as
\begin{equation}
  \sigma_i(t+\delta t) = k \ \text{with prob.} \ \frac{N_k(t)}{N},
\end{equation}
where $N_k(t)$ is the number of agents with opinion $k$ at time $t$.
In order to endow this system with personalized information (PI) we consider
also $N$ PI fields $e_i$, each coupled with
the corresponding standard voter agent $\sigma_i$.
The state of PI fields is a random variable ranging from $1$ to $M$
and which assumes the value $k$ with
probability $P\qua*{e_i(t)=k}$. This probability varies from agent to
agent and over time, depending on the history of the corresponding
voter: the more a voter has chosen a given opinion in the past, the higher
the probability that its corresponding PI field suggests that opinion.
In order to quantify this reinforcement process we generalize the expression
for $P\qua*{e_i(t)=k}$ proposed in \cite{DeMarzo2020} to the case of $M$
distinct opinions, more precisely
\begin{equation}
  P\qua*{e_i(t)=k}=P\qua*{n^{(k)}_i(t)}=\frac{c^{n_i^{(k)}}(t)}{\sum_{j=1}^Mc^{n_i^{(j)}}(t)},
  \label{eq:prob_PI}
\end{equation} 
where $n_i^{(k)}(t)$ is the number of times agent $i$
has chosen (or also confirmed) opinion $k$ up to time
$t$; in the following we will often write $n_i^{(k)}$,
keeping the time dependence implicit. The state of the
system is thus described by $N(M+1)$ variables, namely
$\{(\sigma_i,n_i^{(k)})\}$, and the dynamics takes
place as in the voter model with personalized
information~\cite{DeMarzo2020}. Initially each opinion
$\sigma_i$ is set equal to a random value; at each
time step a given agent $i$ is selected uniformly at random and
with probability $1-\lambda$ it follows the usual
voter dynamics, while with probability $\lambda$ the agent
copies the state $e_i(t)$ of the corresponding PI field.
More explicitly
\begin{equation}
  \sigma_i(t+\delta t)=
  \begin{cases}
    e_i(t) \ \text{with prob. }\lambda\\
    k\ \text{with prob. }(1-\lambda)\frac{N_k}{N}
  \end{cases}
  \label{eq:update_vmpi}
\end{equation}
As in~\cite{DeMarzo2020}, the parameter $\lambda$ sets the strength
of the personalized information with respect to the interaction with
other individuals, while $c$ determines how fast personalized
information adapts to the preferences of agents.

\subsection{Phenomenology of the multistate voter model with personalized information}
\begin{figure}
  \centering
  \includegraphics[width=\columnwidth]{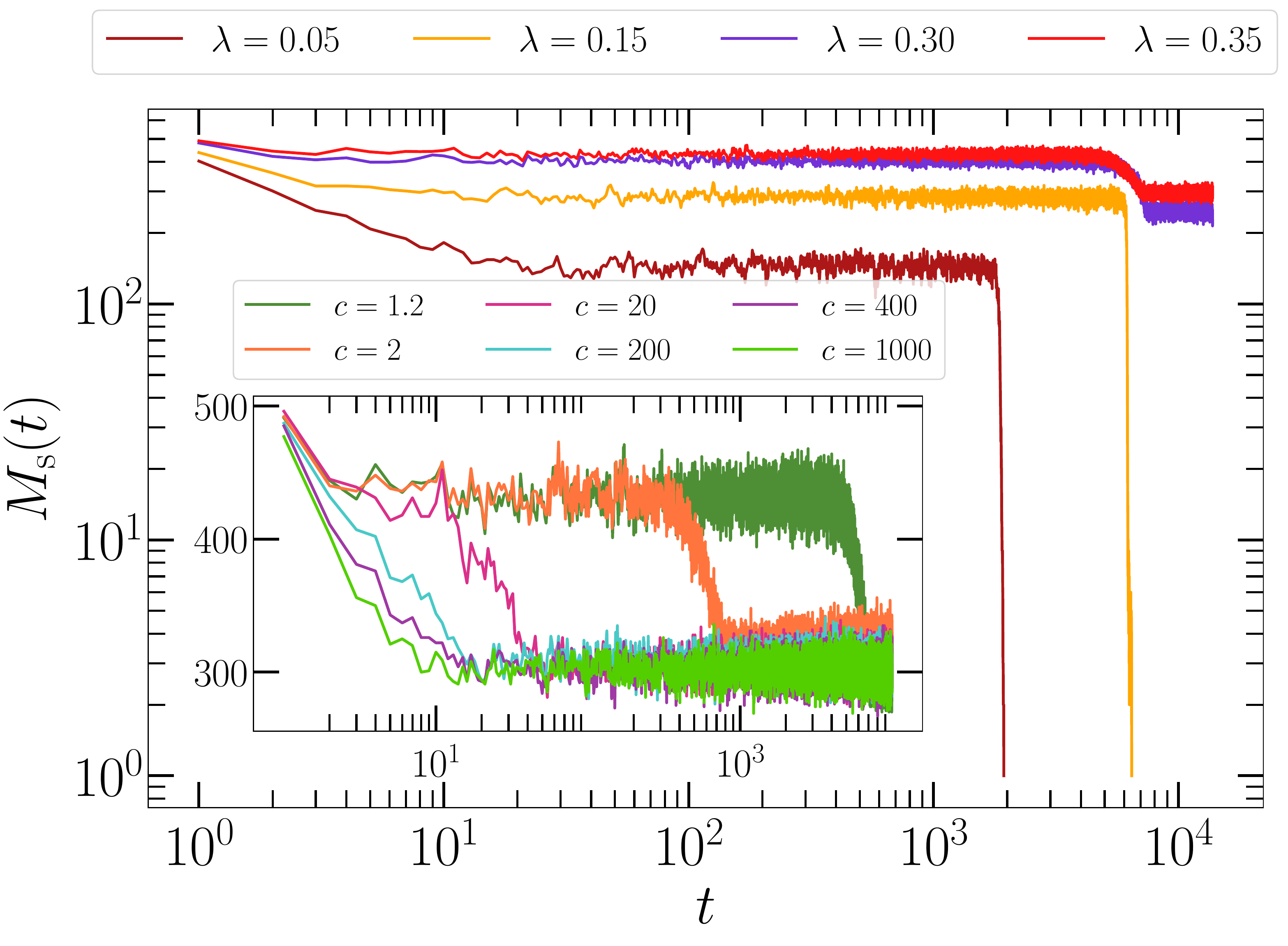}
  \caption{Temporal evolution of the number of surviving
    opinions for $N=1000$, $M=1000$, $c=1.2$ and various
    values of $\lambda$. The inset shows the same plot
    for $\lambda=0.35$ and various values of
    $c$.}\label{fig:small_c}
\end{figure}
Let us illustrate the overall qualitative phenomenlogy of the model,
common for generic values of $M$, $N$ and $c$.
A crucial observable is the number of surviving
opinions $M_s(t)$ as function of time, which plays the role of the order
parameter of the system. In the absence of personalized information (that
is for $\lambda=0$), for a homogeneous initial condition with
$M_s(0)=M$ states, the average value of $M_s(t)$ satisfies
for $t \ll N$~\cite{Starnini2012}
\begin{equation}
  M_s(t)=\frac{M}{1+\frac{M}{N}t}.
  \label{eq:M_s_starnini}
\end{equation}
For times of the order of  $N$, $M_s$ reaches the value $M_s=1$,
implying an ordered configuration (consensus) in which all
agents share the same opinion, as also shown in Fig.~\ref{fig:large_c}a.
Notice also that there is an intrinsic timescale $t_0=N/M$
in Eq.~\eqref{eq:M_s_starnini} such that
$M_s$ remains constant and equal to $M_s(0)$ for times up to $t_0$.
As $\lambda$ increases deviations from
Eq.~\eqref{eq:M_s_starnini} start to appear, but the system keeps
reaching the ordered state after a sufficiently large amount of time, see
Fig.~\ref{fig:large_c}b. The behavior changes as $\lambda$ approaches
the threshold $\lambda_c$; in this case consensus can still be
reached, but in some realizations the system remains trapped in a
stable disordered state (also denoted as polarized state) characterized by the
presence of more than one opinion (Fig.~\ref{fig:large_c}c).
Finally, for $\lambda$ substantially larger
than $\lambda_c$, the system never reaches consensus and the
asymptotic state is always the disordered one (Fig.~\ref{fig:large_c}d).
This qualitative phenomenology is observed
independently of the value of $c$ (provided that $c>1$). This parameter
only determines the possible presence of an initial transient dominated by
the randomness of personalized information in the first time
steps, as shown Fig.~\ref{fig:small_c}. Indeed for $c=1+\delta$ with
$\delta\ll1$ the probability of PI fields, Eq.~\eqref{eq:prob_PI},
can be approximated as
\[
P\qua*{n^{(k)}_i}=\frac{(1+\delta)^{n_i^{(k)}}}{\sum_{j=1}^M(1+\delta)^{n_i^{(j)}}}\approx \frac{1+n_i^{(k)}\delta}{M+\delta\sum_{j=1}^Mn_i^{(j)}}.
\]
Exploiting the fact that $\sum_{j=1}^Mn_i^{(j)}$ is nothing but the
number of times agent $i$ has been updated and that going from $t$ to
$t+1$ each agent is on average updated once we have
$\sum_{j=1}^Mn_i^{(j)}\approx t$ so that
\[
P\qua*{n^{(k)}_i}\approx \frac{1+n_i^{(k)}\delta}{M+t \delta}\approx\frac{1}{M}+\frac{\delta}{M}\ton*{n_i^{(k)}-\frac{t}{M}}.
\]	
Since $0<n_i^{(k)}<t$ this implies 
\[
\frac{1}{M}-\delta\frac{t}{M^2}<P\qua*{n^{(k)}_i}<\frac{1}{M}+\delta\frac{t}{M}\ton*{1-\frac{1}{M}}.
\]
As a consequence $P\qua*{n^{(k)}_i}=\frac{1}{M}+O(\delta t)$ and so
for $\delta$ small the dynamics is initially equal to that of a
multistate voter model with external random field. This produces the
observed transient. A detailed analysis of the multistate voter model in presence of random external information is beyond the scope of this work, the interested reader can find further details in \cite{herrerias2019consensus}.
In order to determine an upper bound of the $c_{th}$ above which no transient is
observed, let us consider the first
update of agent $i$ and let us suppose that opinion $m$ is selected.
As a consequence the probability of the corresponding PI field for
the successive update is
\[
P\qua*{e_i=k}=
\begin{dcases}
  \frac{c}{c+M-1}\ \text{for } k=m\\
  \frac{1}{c+M-1}\ \text{for } k\neq m.
\end{dcases}
\]
If the probability of the PI to be in state $m$ is larger than the
probability of being in any other state, a reinforcing loop gets
established and the PI gets more and more polarized along opinion $m$. 
This implies that a sufficient condition for observing a polarized PI
already after the first step is
\[
P\qua*{e_i(\delta t)=m}>\sum_{k\neq m}P\qua*{e_i(\delta t)=k} ,
\]
which implies
\[
\frac{c}{c+M-1}>\frac{M-1}{c+M-1}
\]
yielding a threshold value
\be
c_{th}=M-1.
\ee

Fig.~\ref{fig:small_c} shows that this estimate is an upper bound of the real $c_{th}$. The duration of the transient governed by a random external field gets
shorter as $c$ is increased and it is completely absent for $c \approx c_{th}=999$.
Since $c$ plays only a marginal role, in the rest of the paper we focus
on the case $c > c_{th}$ so as to remove the initial transient.

\begin{figure*}
  \centering
%  \begin{subfigure}[b]{0.32\textwidth}
%    \centering
%    \includegraphics[width=\textwidth]{fig_4a_Ps}
%    \caption{}
%    \label{fig:Ps}
%  \end{subfigure}
%  \hfill
  \begin{subfigure}[b]{0.49\textwidth}  
    \centering 
    \includegraphics[width=\textwidth]{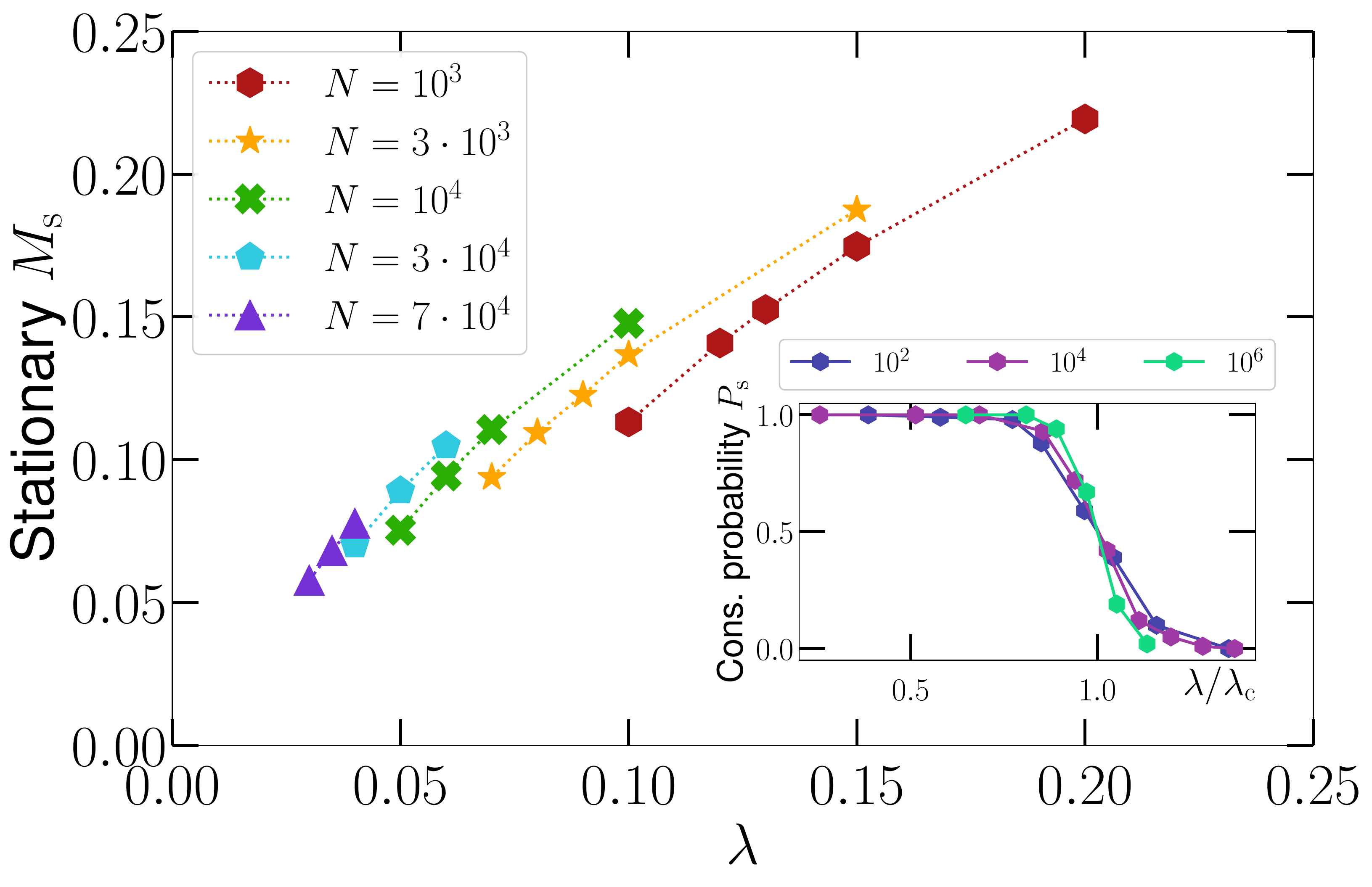}
    \caption{}    
    \label{fig:Ms_vs_lambda}
  \end{subfigure}
  \hfill
  \begin{subfigure}[b]{0.49\textwidth}
    \centering
    \includegraphics[width=\textwidth]{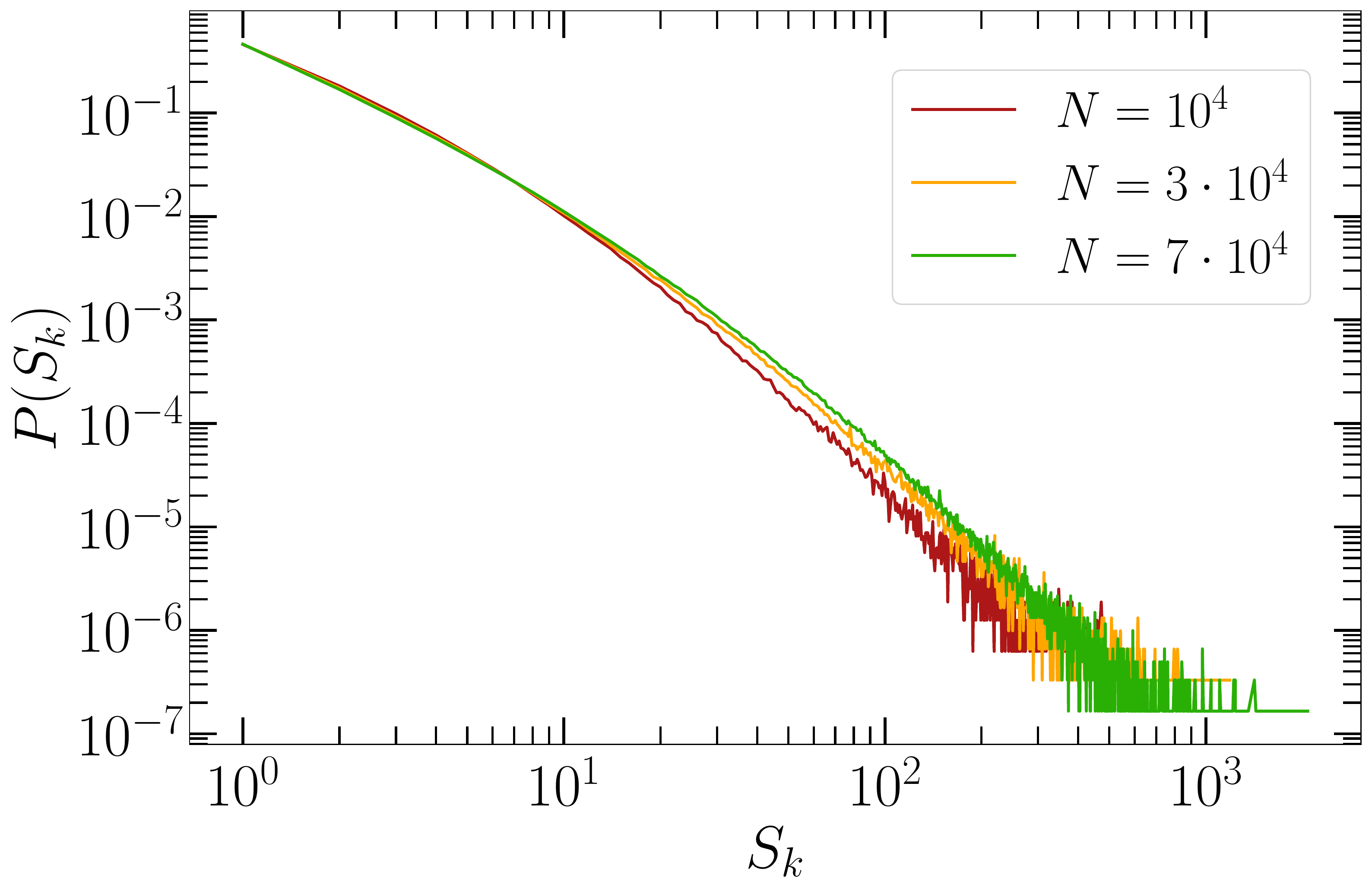}
    \caption{}     
    \label{fig:power_law}
  \end{subfigure}

  \caption{a) Main: Stationary value of the fraction $M_s$ of
    opinions, for $M=N$, averaged only over surviving
    runs. Inset: Probability
    that consensus is reached as a function of $\lambda$ for $N=10^2$, $10^4$, $10^6$ and $M=20$. As $N$ increases the transition between consensus and polarization becomes sharper and sharper. b) Probability distribution of the number of agents $S_k$
    polarized along opinion $k$ at criticality for $N=M$. This has been obtained by letting the system evolve toward the stationary state and then performing a binning over the $S_k$ so to obtain their histogram. Note that such an histogram corresponds to a single realization of the system.} 
  \label{fig:continuous_transition}
\end{figure*}

Finally, we illustrate the nature of the transition observed
as $\lambda$ is varied.
As evident from Fig.~\ref{fig:large_c}, around the transition
different realizations of the process lead to different outcomes:
either consensus or a stationary state.
The transition is characterized by the variation, as a function
of $\lambda$, of the fraction $P_s$ of runs reaching a
stationary state (Fig.~\ref{fig:Ms_vs_lambda},inset). As it is possible to see, the larger is $N$, the sharper the transition becomes. Moreover, since for large $N$ the critical threshold $\lambda_c$ goes to zero (see Subsec.~\ref{subsec:transition_point} and figures therein) and the transition gets very sharp, this implies that in large systems even an infinitesimal amount of personalized information is sufficient to make the reaching of consensus impossible.
The main panel of Fig.~\ref{fig:Ms_vs_lambda} displays how the fraction
of surviving opinions in the stationary state $M_s(t\to\infty)/M=M_s^{\infty}/M$,
averaged only over surviving runs, varies as a function of $\lambda$.
The quantity $M_s^{\infty}/M$ grows in a continuous fashion, starting
from a finite value decreasing with $N$.
This suggests that in the large $N$ limit the transition is continuous,
as also confirmed by inspecting the distribution of $S_k(t\to\infty)=S_k^{\infty}$, 
the number of agents polarized along opinion $k$ in the stationary
state. When $c>c_{th}$ the number of polarized agents $S_k$ along opinion $k$ is given by the number of agents such that $n_i^{(k)}>n_i^{(j)}$ for any $j\neq k$, since, in this case, the personalized information suggest the most chosen opinion. Indeed, by looking at Fig.~\ref{fig:power_law}
it is clear that such a distribution decays as a power-law for
$\lambda=\lambda_c$, with a nontrivial exponent approximately equal to 2.8 (for $N=M$).

\section{Analytical approach}
As shown above, the phenomenology of the model is only
marginally influenced by the parameter $c$.
Therefore consider the case $c\gg c_{th}$
which can be more easily handled analytically.
Indeed in such a situation the probability of the
PI field $e_i$ is strongly peaked on the opinion more frequently
held by agent $i$ and
Eq.~\eqref{eq:prob_PI} reduces to
\begin{equation}
  P\qua*{e_i=k}=\delta_{k,m},
  \label{Prob_PI_large_c}
\end{equation}
where $m$ is the opinion which satisfies
\[
m=\underset{k}{\text{argmax}}\qua*{n_i^{(k)}}.
\]
In other words the PI suggests only the favorite opinion
in the past.
Our goal is to derive, under the assumption of large $c$,
analytical estimates of the critical value $\lambda_c$
for generic large values of $N$ and $M$.

\subsection{Stability of polarized states}
\begin{figure*}
	\centering
	\begin{subfigure}{.45\textwidth}
		\includegraphics[width=\textwidth]{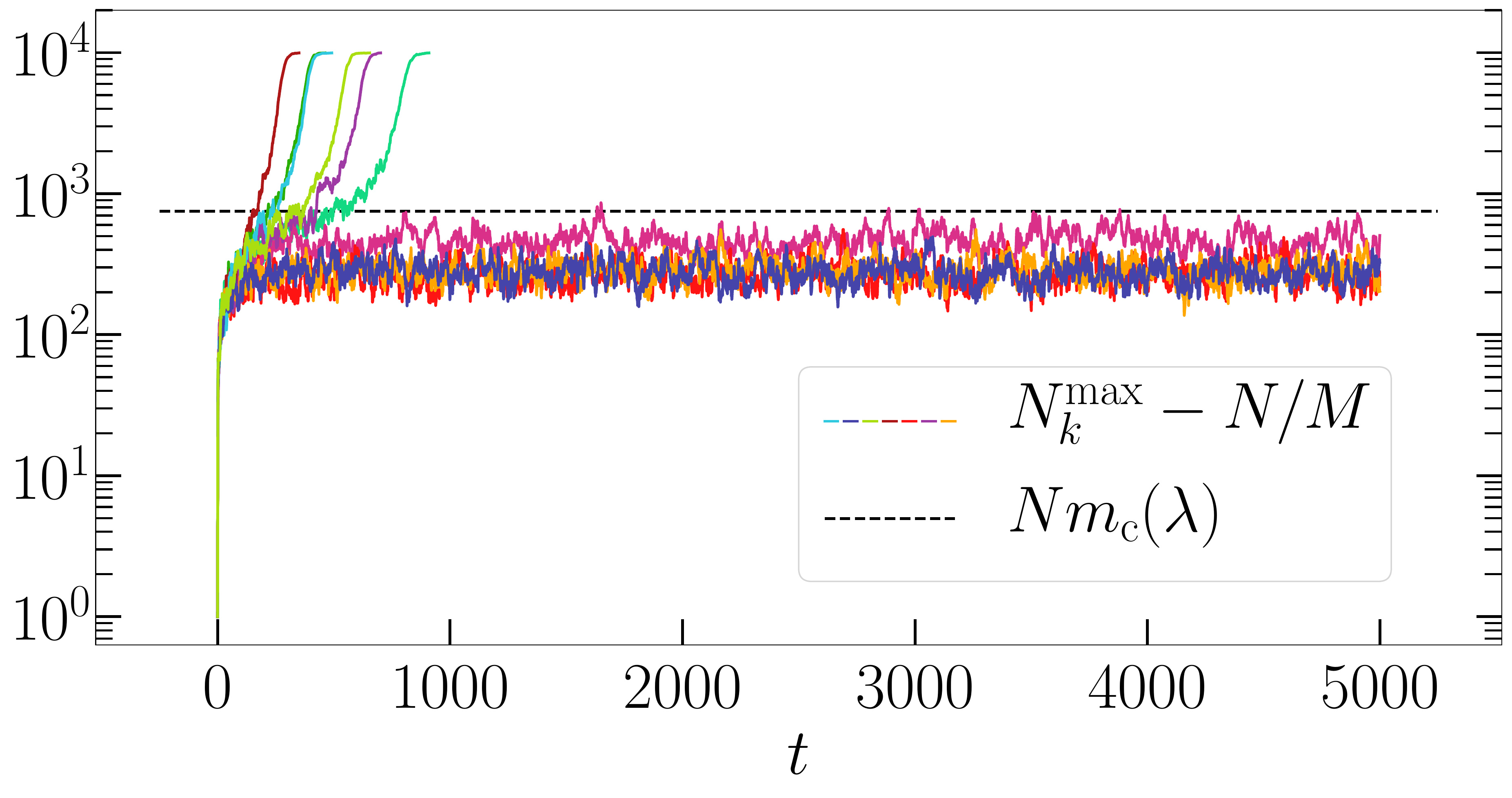}
		\caption{}
	\end{subfigure}\hfill%
	\begin{subfigure}{.45\textwidth}
		\includegraphics[width=\textwidth]{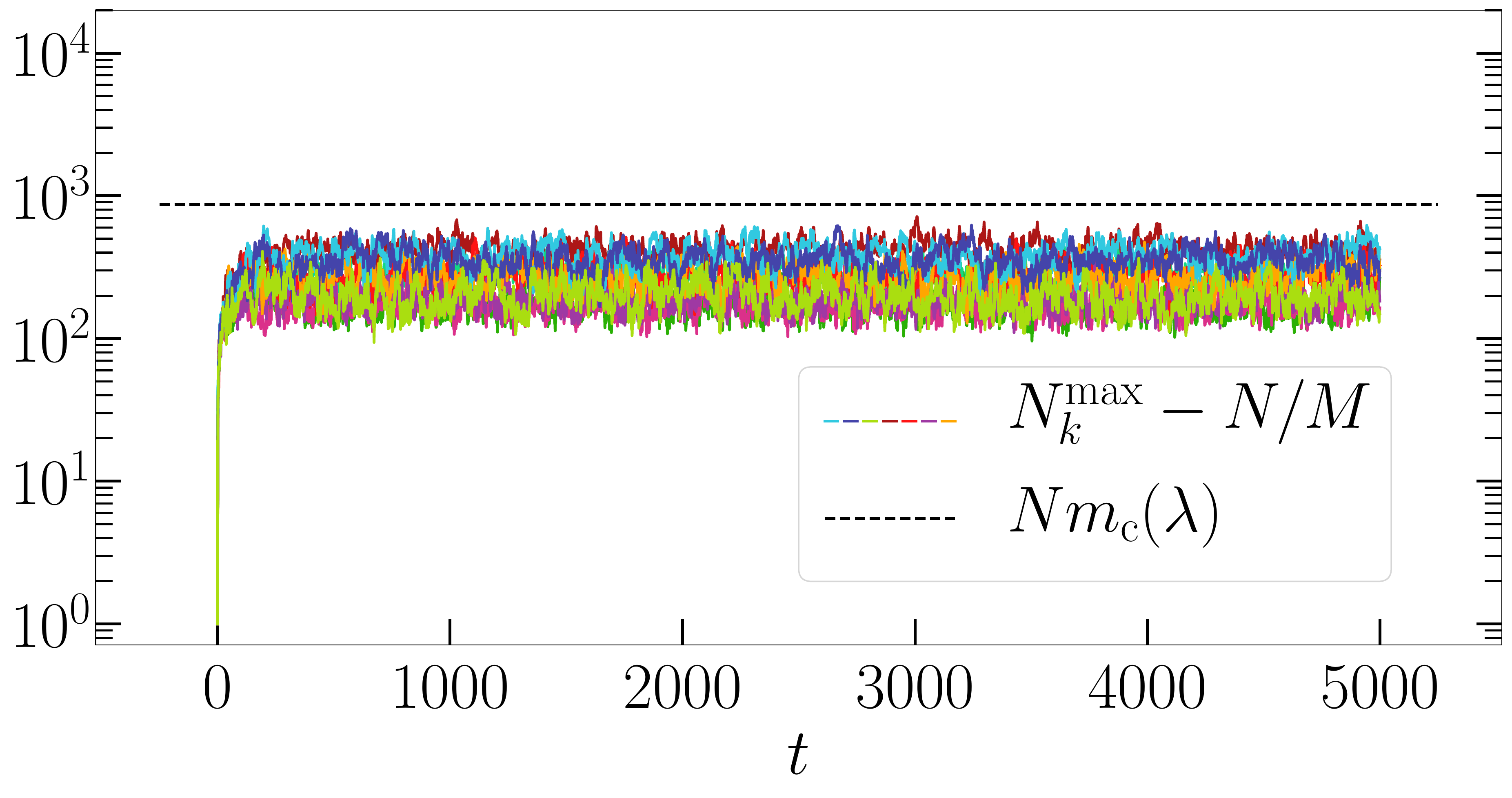}
		\caption{}
	\end{subfigure}
	\caption{Temporal evolution of $N_{\max}-N/M$, compared with the critical value $N m_c(\lambda)$
          for $N=10^4$, $M=200$: a) $\lambda=0.07$; b) $\lambda=0.08$. Different colors correspond to different simulations.}
        \label{fig:m_c}
\end{figure*}
Let us consider an agent $i$ in state $\sigma_i$ and let
us assume that its PI is polarized along opinion $m$. Combining
Eqs.~\eqref{eq:update_vmpi} and \eqref{Prob_PI_large_c} we can write
the transition probability of this agent~\footnote{Note that these
transition probabilities do not depend on the state $\sigma_i$ of
the agent} as
\be
W\ton*{\sigma_i \to k}=
\begin{dcases}
  \frac{1}{N}\qua*{\lambda+(1-\lambda)\frac{N_k}{N}} &\text{ for }k=m\\
  \frac{1}{N}(1-\lambda)\frac{N_k}{N}&\text{ for }k\neq m.
\end{dcases}				
\label{W}
\ee
For the overall stability of the polarized state, the PI field $e_i$ should
remain polarized on opinion $m$ so as to keep the agent we are
considering fixed (on average) on this same opinion. This requires
the transition probability to opinion $m$ to be larger than the
transition probability to any other opinion $l$, otherwise $n_i^{(l)}$
would grow faster than $n_i^{(m)}$ and the PI would eventually
depolarize and then polarize along opinion $l$.
As a consequence, in order for the polarized state to be stable it must be
\[
W\ton*{\sigma_i \to m}>W\ton*{\sigma_i \to l}~~~~~~\forall l \neq m,
\]
which yields, using Eq.~(\ref{W})
\begin{equation}
  \frac{N_m-N_l}{N}>-\frac{\lambda}{1-\lambda}=-m_c(\lambda).
  \label{eq:stability_preliminary}
\end{equation}
Here, in analogy with~\cite{DeMarzo2020}, we defined the critical
magnetization $m_c(\lambda)$ as
\begin{equation}
  m_c(\lambda) = \frac{\lambda}{1-\lambda}.
  \label{eq:m_c}
\end{equation}
We can then repeat this same reasoning but considering an agent $j$
whose PI field $e_j$ is polarized along opinion $l$. This leads to
\[
\frac{N_m-N_l}{N}<m_c(\lambda).
\]
Combining this expression with
Eq.~\eqref{eq:stability_preliminary} we obtain the following necessary
condition for a polarized state to be stable
\begin{equation}
  \left | \frac{N_m-N_l}{N} \right|<m_c(\lambda) ~~~~\forall(l, m).
  \label{eq:condition_stability_N}
\end{equation}

By introducing the partial
magnetization $m_k$ along opinion $k$,
we can also rewrite this constraint as

\[
m_k=\frac{N_k-\sum_{l\neq k}N_k}{N}=\frac{2N_k-N}{N}.
\]
In this way Eq.~\eqref{eq:condition_stability_N} becomes
\begin{equation}
  \left|\frac{m_m-m_l}{2}\right|<m_c.
  \label{eq:condition_stability_m}
\end{equation}
The conclusion if the difference between any two magnetizations is smaller than the threshold $m_c$ then the polarized state is stable. Note, however, that this condition is very strict and a polarized state can be stable on average even if there are some opinions held by a small number of agents not fulfilling it. Indeed in such a situation these opinions will be absorbed by the others, but still consensus will not be reached. As a consequence what really matters is that \eqref{eq:condition_stability_N} is fulfilled when considering the most common opinion whose size is $N_{max}$ and the average opinion size given by $N/M$. The condition ensuring the system not to reach consensus is thus
\begin{equation}
  N_{max}-\frac{N}{M} <Nm_c(\lambda) .
  \label{eq:condition_stability_N_average}
\end{equation}

Fig.~\ref{fig:m_c} checks and confirms the validity of this argument in numerical simulations.
Note that $m_c$ is larger than 1 for $\lambda > 1/2$ and so, as in
the voter model with personalized information, above $\lambda=1/2$ an
opinion can survive even with only a single agent supporting it.

\subsection{Polarization time}
\label{subsec:polarization_time}
\begin{figure}
%  \centering
    \includegraphics[width=\columnwidth]{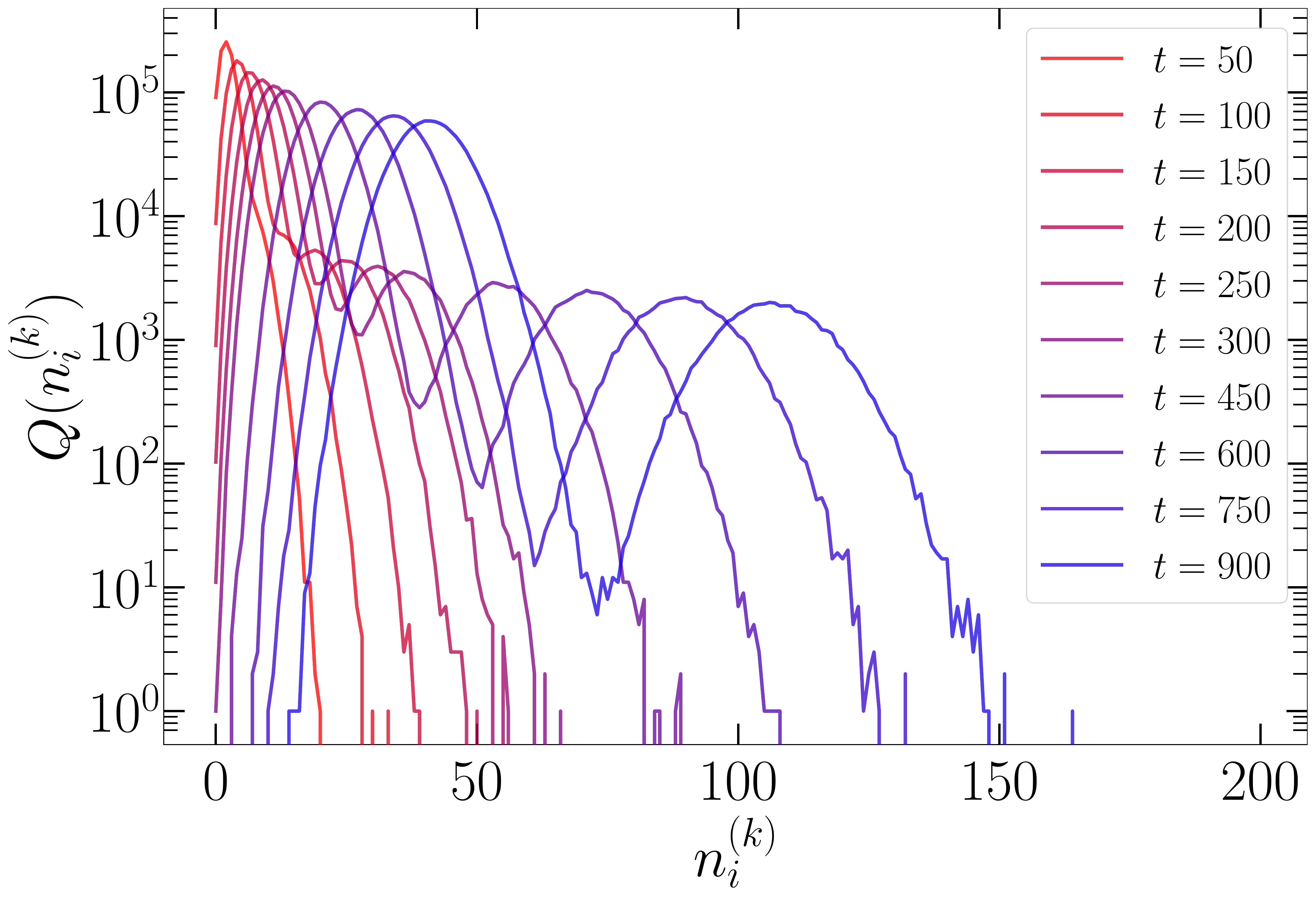}
    \caption{Temporal evolution of the $Q(n_i^{(k)})$ distribution as
      a function of time.  Data are for $c=400,$ $N=10^6$, $M=20$,
      $\lambda=0.07$.  The component with high mean value represents
      agents polarized along opinion $k$; the component with low mean
      value represents agents unpolarized or polarized along other
      opinions.}
  \label{fig:splitting}
%  \hfill
%  \begin{subfigure}[b]{0.3\textwidth}
%    \centering
%    \includegraphics[width=0.9\textwidth]{Screenshot_20210917_141521.png}
%  \caption{Splitting2}
%  \label{fig:splitting2}
%  \end{subfigure}
%  \hfill
%  \begin{subfigure}[b]{0.3\textwidth}  
%    \centering 
%    \includegraphics[width=0.9\textwidth]{Screenshot_20210917_141624.png}
%    \caption{Splitting3}    
%    \label{Splitting3}
%  \end{subfigure}
\end{figure}
In deriving Eq.~\eqref{eq:condition_stability_m} we assumed all PI
fields to be polarized on a certain opinion, meaning that for any $i$
there is only one $k$ such that $n_i^{(k)}$ is maximum. However
initially $n_i^{(k)}=0$ for all pairs $(i, k)$ and the randomness of voter
updates implies the occurrence of several ties between the various
$n_i^{(k)}$.
The assumption that PI fields are polarized starts to be valid
only after a polarization time $t^*>0$. In order to compute this
time we write down the master equation for the distribution of
the $n_i^{(k)}$ values, $Q\ton{n_i^{(k)}}$, that reads
\begin{equation}
  \frac{dQ\ton*{n_i^{(k)}}}{dt}=NR\ton*{n_i^{(k)}-1}Q\ton*{n_i^{(k)}-1}-NR\ton*{n_i^{(k)}}Q\ton*{n_i^{(k)}},
  \label{eq:master_equation_n_general}
\end{equation}
where we introduced the single transition probability 
\begin{align}
  NR\ton*{n_i^{(k)}}&=\ton*{1-\lambda}\frac{N_k}{N}+\lambda P\ton*{e_i=k} \nonumber \\ %\approx R\ton*{n_i^{(k)}}=\nonumber\\
  &\approx \ton*{1-\lambda}\frac{1}{M_s(t)}+\lambda P\ton*{e_i=k}.
  \label{eq:transition_R_general}
\end{align}
Here we assumed that all surviving opinions share
approximately the same number of agents and so we can make the
approximation ${N_k}/{N}\approx{1}/{M_s(t)}$. Now, considering
opinion $k$, only a fraction $S_k/N$ of the agents is polarized along
this opinion and thus we assume that the distribution
$Q\ton{n_i^{(k)}}$ is formed by two normalized components $Q_1$ and
$Q_2$ corresponding to the two types of agents: those polarized on
opinion $k$ (component $Q_2$) and those polarized on any other
opinion or not polarized at all (component $Q_1$).
We can then write $Q$ as a bimodal distribution
\[
Q\ton*{n_i^{(k)}}=\ton*{1-\frac{S_k}{N}}Q_1\ton*{n_i^{(k)}}+\frac{S_k}{N}Q_2\ton*{n_i^{(k)}}.
\]
Both components evolve according to the same general master equation
Eq.~\eqref{eq:master_equation_n_general}, but the transition rates are
different. For what concerns the polarized component $Q_2$ we have
\[ 
NR\ton*{n_i^{(k)}}=\ton*{1-\lambda}\frac{1}{M_s(t)}+\lambda 
\] 
and using this expression and Eq.~\eqref{eq:master_equation_n_general}
we can compute both the mean value $n_2$ and the variance $\sigma_2^2$
of the distribution $Q_2$ for polarized agents, finding
\begin{align}
  n_2 = \sigma_2^2 = \int_0^t dt' \ton*{\frac{1-\lambda}{M_s(t')}+\lambda},
  \label{eq:mean_variance_2}
\end{align}
Analogously, the transition rate of the component $Q_1$, corresponding
to agents not polarized on opinion $k$, is
\[
NR\ton*{n_i^{(k)}}=\ton*{1-\lambda}\frac{1}{M_s(t)}
\]
and so the mean value $n_1$ and the variance $\sigma_1^2$ are 
\begin{align}
  n_1 = \sigma_1^2 = \int_0^t dt' \ton*{\frac{1-\lambda}{M_s(t')}}.
  \label{eq:mean_variance_1}
\end{align}
Detailed computations are reported in
Appendix~\ref{apx:Master equation, transition rates and moments}.

Looking at Eqs.~\eqref{eq:mean_variance_2}
and \eqref{eq:mean_variance_1} it is clear that initially the two
components are indistinguishable since they both have null mean and
variance. However, having different velocities they
tend to separate over time.
This behavior is shown in Fig.~\ref{fig:splitting}.
Until $Q_1$ and $Q_2$ are superposed there is
no real distinction between polarized and unpolarized agents, as all agents
can change their preferred opinion in one or few voter updates.
Consequently we identify the polarization time $t^*$ introduced above
as the time when the two components split for the first time.
This can be readily determined by imposing the distance between the
two peaks to be equal to their widths, that is
\[
n_2(t^*)-n_1(t^*)=2\qua*{\sqrt{\sigma_1^2(t^*)}+\sqrt{\sigma_2^2(t^*)}},
\]
where the factor $2$ is somewhat arbitrary and does not influence the
scaling of $t^*$. Substituting Eqs.~\eqref{eq:mean_variance_2} and
\eqref{eq:mean_variance_1} into this last expression we come up with
an implicit integral expression for the polarization time
$t^*$
\begin{equation}
  t^*=\frac{2}{\lambda}\gra*{\sqrt{\int_0^{t^*}\qua*{\frac{1-\lambda}{M_s(t)}+\lambda}dt'}+\sqrt{\int_0^{t^*}\frac{1-\lambda}{M_s(t)}dt'}}
  \label{eq:polarization_time}
\end{equation}
%In order to solve this equation and obtain explicit estimates for
%$t^*$ one has to make an assumption on the form of $M_s(t)$, as
%is done in the next subsection.

\subsection{The transition point}
\label{subsec:transition_point}

Once the polarization time has been obtained we can turn to
the determination of the critical parameter $\lambda_c$.
The idea is the following: In the initial stage of the dynamics,
the numbers of agents holding the different opinions, $N_l$, tend
to fluctuate and their differences tend to grow over time.
If this growth is slow enough, at the polarization time $t^*$
the condition for the stability of polarized states,
Eq.~\eqref{eq:condition_stability_N_average} is satisfied. In such a case the disordered state with multiple
coexisting opinions is stable and persists forever.
Conversely, if at $t^*$ Eq.~\eqref{eq:condition_stability_N_average}
is violated, the system will eventually evolve toward the consensus state. The condition for $\lambda_c$ thus reads
\begin{equation}
  x_{\max}(t^*)=N_{\max}(t^*)-\frac{N}{M}=N\frac{\lambda_c}{1-\lambda_c}.
  \label{eq:condition_lambda_c_begin}
\end{equation}
By means of numerical simulations and a simple scaling argument
(see Appendix~\ref{apx:Scaling of x_max}) we determine that for $M \ll N$ $x_{\max}(t)$ scales as
\begin{equation}
  x_{\max}(t) \sim \gamma\frac{N^{1/2}}{M}t^{3/2},
  \label{eq:x_max(t)}
\end{equation}
where $\gamma\approx1/4$ is a numerical constant. We also derived an
equation for the time growth of $N_k-N/M$ and we checked that the
temporal scaling of $x_{\max}(t)$ is consistent with this expression,
see Appendix~\ref{apx:evolution_xk_yk}.  Replacing this last
expression into Eq.~\eqref{eq:condition_lambda_c_begin} we obtain an
equation for $\lambda_c$
\[
\gamma\frac{N^{1/2}}{M}{t^*}^{3/2}=N\frac{\lambda_c}{1-\lambda_c},
\]
implying
\begin{equation}
  \lambda_c = \frac{\gamma \frac{N^{1/2}}{M}\qua*{t^*(\lambda_c)}^{3/2}}{N+\gamma \frac{N^{1/2}}{M}\qua*{t^*(\lambda_c)}^{3/2}}.
  \label{eq:condition_lambda_c_final}
\end{equation}

Eq.~\eqref{eq:condition_lambda_c_final}, together with
Eq.~\eqref{eq:polarization_time} for the polarization time,
provide a closed system of equations in $\lambda_c$ and $t^*$
\begin{equation}
	\begin{dcases}
    t^*=\frac{2}{\lambda_c}\gra*{\sqrt{\int_0^{t^*}\qua*{\frac{1-\lambda_c}{M_s(t')}+\lambda_c}dt'}+\sqrt{\int_0^{t^*}\frac{1-\lambda_c}{M_s(t')}dt'}}\\
    \lambda_c = \frac{\gamma \frac{N^{1/2}}{M} (t^*)^{3/2}}{N+\gamma \frac{N^{1/2}}{M} (t^*)^{3/2}}.
  \end{dcases}
  \label{eq:system}
\end{equation}

%However we are still left with $M_s(t)$ in the first equation, so

In order to solve this system we need an explicit expression for the
number $M_s(t)$ of surviving opinions.
We consider two possible assumptions, which are expected to be fairly
accurate for different values of $M$ and $N$.
Here we limit ourselves to report the
main results, the interested reader can find detailed
calculations in Appendix~\ref{apx:scaling_regimes}.

\subsection{$\mathbf{M \ll N}$}

%\item $\mathbf{M \ll N}$ \\
  If the number of opinions in much smaller
  than the number of agents, each opinion is initially shared by
  a large number of individuals. As a consequence during a first time
  interval no opinion disappears and it is reasonable to make the
  approximation
  \[
  M_s(t)=M 
  \]
  In this way Eq.~\eqref{eq:system} can be rewritten as
  \begin{equation}
  \begin{dcases}
    t^*=\frac{4}{\lambda_c^2}\qua*{\sqrt{\frac{1-\lambda_c}{M}+\lambda_c}+\sqrt{\frac{1-\lambda_c}{M}}}^2\\
    \lambda_c = \frac{\gamma \frac{N^{1/2}}{M}(t^*)^{3/2}}{N+\gamma \frac{N^{1/2}}{M}(t^*)^{3/2}}.
  \end{dcases}
  \label{eq:system_M<<N}
  \end{equation}
  Depending on how large $M$ is with respect to $N$,
  the solutions of this system scale in different ways.

  \begin{itemize}
  \item $\mathbf{M \ll N^{1/3}}$\\
    \begin{equation}
      \begin{cases}
	t^* \approx 4(MN)^{1/4}\\
	\lambda_c \approx 2\ton*{M^5N}^{-1/8}.
      \end{cases}
      \label{eq:M<<N^1/3}
    \end{equation}
  \item $\mathbf{N^{1/3}\ll M\ll N}$\\
    \begin{equation}
      \begin{cases}
	t^* \approx 2^{8/5}\ton*{NM^2}^{1/5}\\
	\lambda_c \approx 2^{2/5}\ton*{NM^2}^{-1/5}.
      \end{cases}
      \label{eq:M<<N}
    \end{equation}
  \end{itemize}
  
%\item $\mathbf{M=N}$ \\
\subsection{$\mathbf{M=N}$}
If $M=N$ each agent has initially a different
opinion. Hence, even during the first time steps, some opinions
disappear by chance.
In this case we can approximate the surviving opinions
$M_s(t)$ with the expression valid for the simple multistate voter
model, Eq.~\eqref{eq:M_s_starnini}, that is for $M=N$,
\begin{equation}
  M_s(t)=\frac{N}{1+t}.
  \label{Mstarnini}
\end{equation}
Substituting this expression in Eq.~\eqref{eq:system} and imposing
$M=N\gg 1$, $\lambda_c\ll 1$ we obtain the following system
\begin{equation}
  \begin{dcases}
    t^*=\frac{32N^{1/2}+8\lambda_cN}{2\ton*{\lambda_c^2N-16}}\\
    \lambda_c = \frac{\gamma N^{-1/2}\qua*{t^*(\lambda_c)}^{3/2}}{N+\gamma N^{-1/2}\qua*{t^*(\lambda_c)}^{3/2}},
  \end{dcases}
  \label{eq:system_M=N}
  \end{equation}
whose solution scales as
\begin{equation}
  \begin{cases}
    t^* \approx 2^{4/3}N^{2/3}\\
    \lambda_c \approx 4N^{-1/2}
  \end{cases}
  \label{eq:M=N}
\end{equation}
%\end{enumerate}

\begin{figure*}
  \centering
  \begin{subfigure}[t]{.45\linewidth}
    \caption{$M=20 \ll N$.}
    \includegraphics[width=\textwidth]{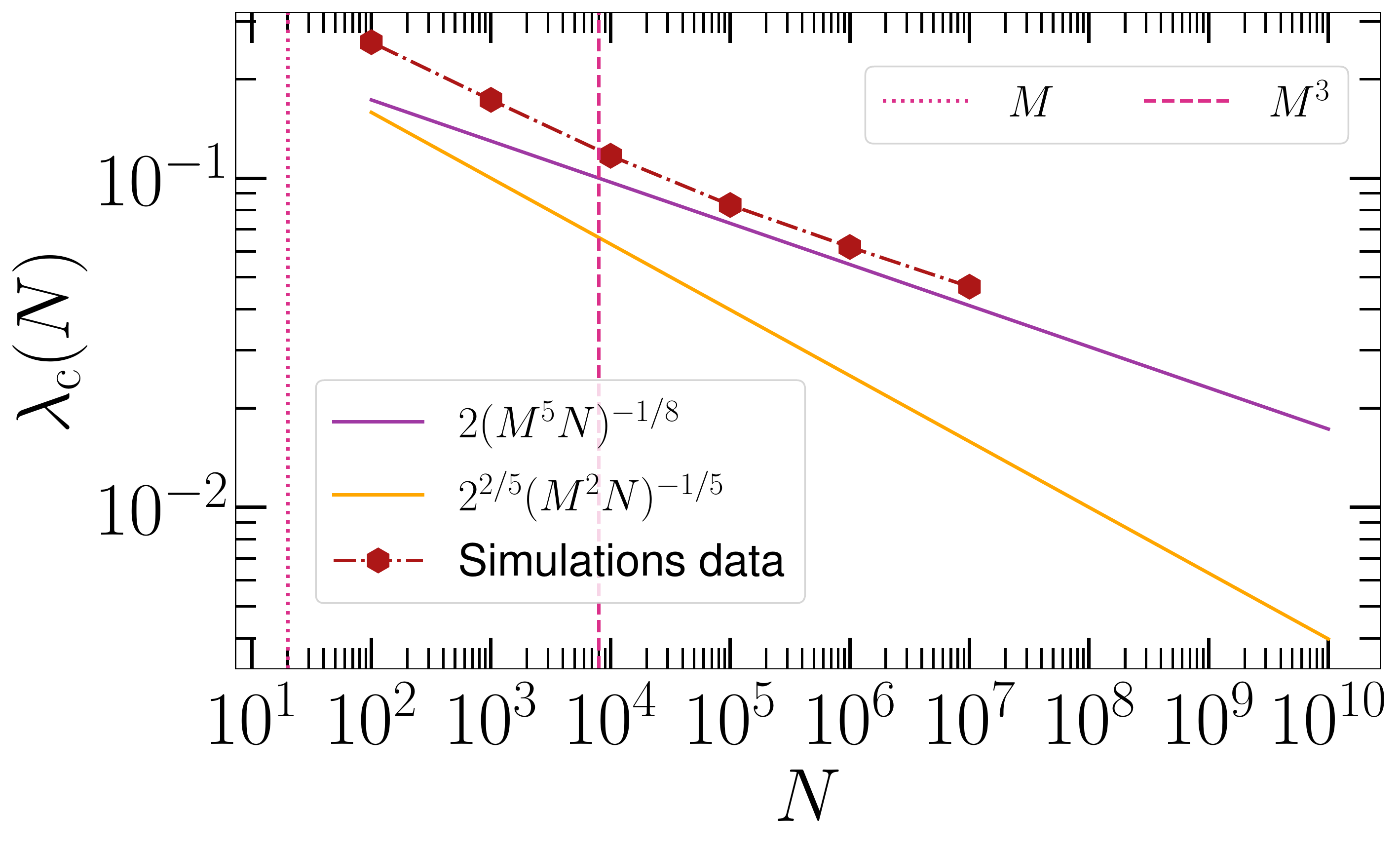}
    \label{fig:theory_vs_simulations}
  \end{subfigure}\hfill%
  \begin{subfigure}[t]{.45\linewidth}
    \caption{$M=200 \ll N$.}
    \includegraphics[width=\textwidth]{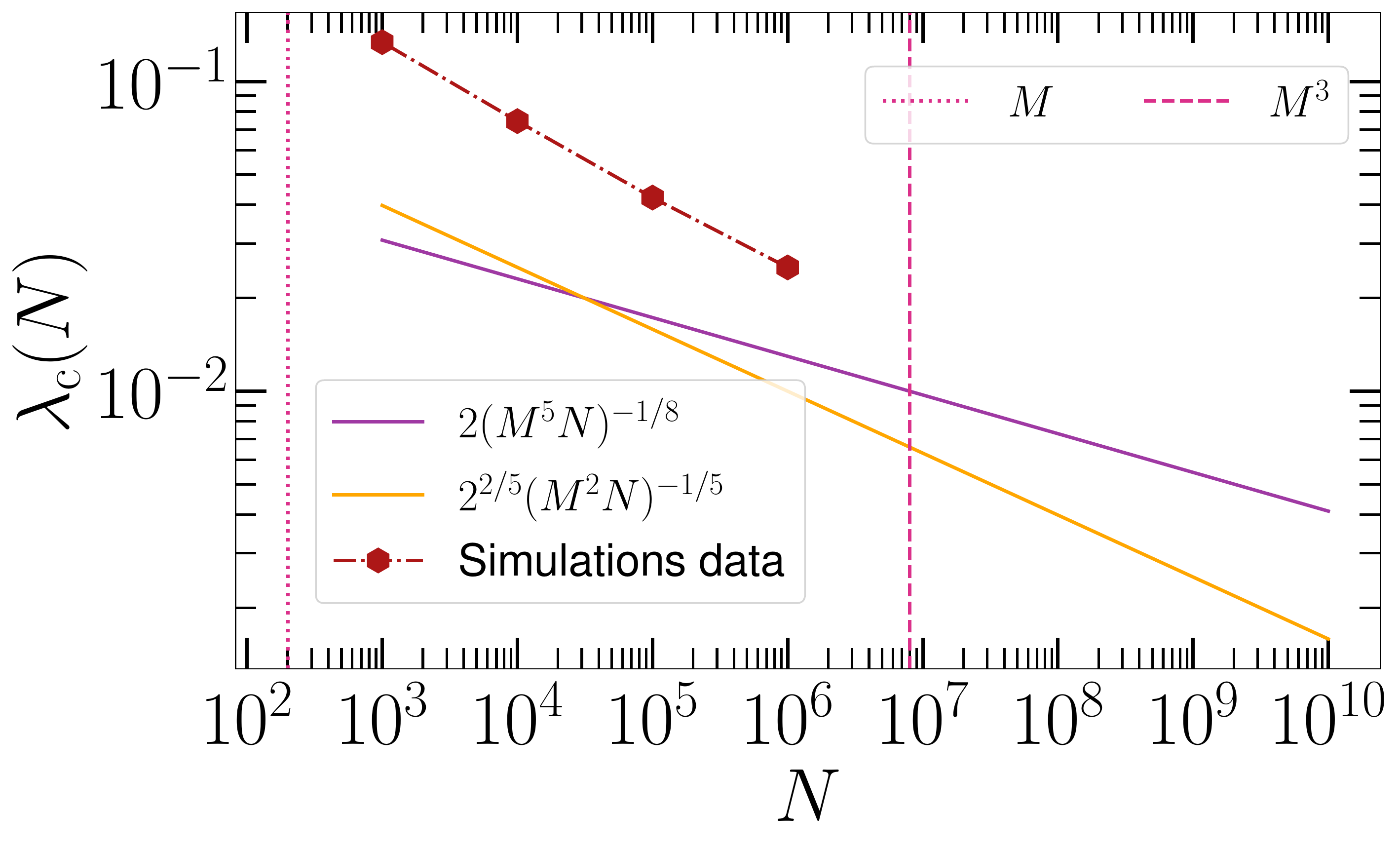}
    \label{fig:theory_vs_simulations2}
  \end{subfigure}\hfill%
  \begin{subfigure}[t]{.45\linewidth}
    \caption{$M=N$.}
    \includegraphics[width=\textwidth]{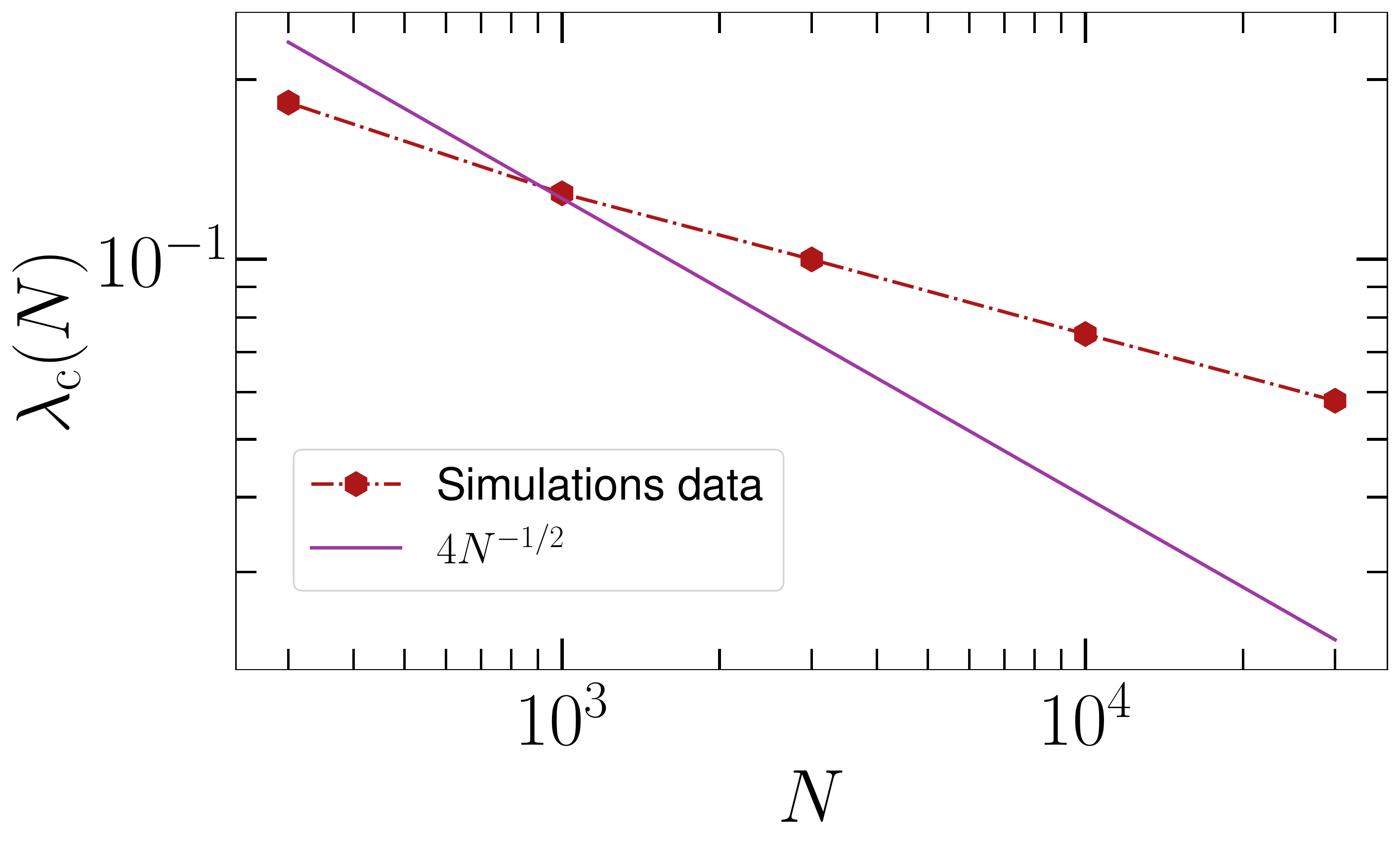}
      \label{fig:theory_vs_simulations3}
  \end{subfigure}
  \caption{Comparison of analytical predictions for the
    threshold $\lambda_c$ and numerical estimates of it}
  \label{}
\end{figure*}

\section{Numerical simulations}

We now compare theoretical predictions with numerical simulations.
We iterate MVM dynamics for various values of $M$ and $N$ up to $t=2000$
and we determine the fraction of times $P_s(\lambda)$ the system
has reached consensus.
We take as numerical estimate of $\lambda_c$ the values such that
$P_s(\lambda_c)=1/2$.

For fixed value of $M$ and increasing size $N$, the analytical
results predict initially a scaling regime given by Eq.~\eqref{eq:M<<N}
as long as $N$ is much larger than $M$ but much smaller than $M^3$,
followed by a scaling given by Eq.~\eqref{eq:M<<N^1/3}.
For $M=20$, the first regime spans a short interval of $N$ values.
In Fig.~\ref{fig:theory_vs_simulations}
we see only a hint of the associated scaling, while for larger values of $N$
the agreement between theory and numerics is very good.
For $M=200$ instead (see Fig.~\ref{fig:theory_vs_simulations2}), the first regime
extends to much larger values of $N$, so that feasible values of $N$ lie only
in this regime.
In this case $\lambda_c$ nicely scales as $N^{-1/5}$, as predicted by
Eq.~\eqref{eq:M<<N}, although the prefactor is not predicted exactly.

For the case $M=N$ instead, Figure~\ref{fig:theory_vs_simulations3}
show that the theory is not able to catch the correct scaling.
This mismatch has various potential origins.
Indeed using Eq.~\eqref{Mstarnini} for the temporal dependence of $M_s(t)$
is a quite rough approximation, as the number of different opinions
actually decays much more slowly over time. But also the use
of Eq.~\eqref{eq:condition_lambda_c_begin} for $x_{\max}(t)$ is not warranted
for $M=N$. A deeper understanding of the phenomenology of the case $M=N$
remains an interesting open question.

\section{Conclusions}

In this paper we have introduced and analyzed a multistate voter model
where the coupling with an external history-dependent individual field
mimics the effect of personalized recommendation algorithms in online
social media.  A population of agents, initially having different
opinions, reaches consensus on a single opinion or remains polarized
on multiple different opinions, depending on the strength of
personalized information.  The phenomenology is governed by the
competition between the fluctuations induced by voter dynamics and the
tendency, due to personalized information, to bind agents to the
opinion they adopted most frequently in the past.  By means of
arguments based on this physical picture, we estimated analytically
the critical threshold between the two regimes, obtaining a reasonable
agreement with simulations for $M \ll N$. Conversely, the dynamics
when the initial number of opinions is comparable with the number of
agents seems to elude our approach.

From a more general point of view, our study indicates how difficult
reaching consensus is, in the presence of personalized
recommendations.  For any number $M$ of initial opinions the threshold
$\lambda_c$ tends to vanish when the number of agents $N$
diverge. This means that no matter how weak is the coupling with the
personalized information, if the system is large enough polarization
unavoidably arises.  In this respect, note that voter dynamics is
extremely favorable to the establishment of consensus: for any
interaction pattern, consensus is necessarily reached for any finite
number of interacting individuals.  The addition of a personalized
recommendation completely changes this picture, at least for large
systems.  For a finite number of interacting agents instead consensus
is still reached if the strength of personalized information is small
enough. The transition between the two regimes exhibits nontrivial
features that may be the focus of future activity along with
generalizations to include nontrivial interaction patterns or a
different functional dependence between the probability distribution
of the personalized information and the number of times an opinion has
been selected in the past.

A natural extension of this work would be to analyze more realistic
recommendation algorithms. With respect to this, a promising
possibility is to consider collaborative filtering algorithms
\cite{schafer2007collaborative}, which are based on similarities among
the history of different users or opinions instead of considering only
the history of the user itself. This typology of algorithm is used in
real life applications \cite{linden2003amazon, smith2017two}, but
still it is sufficiently simple to try an analytical study of its
effects on opinion dynamics models.

\bibliography{\jobname}

%merlin.mbs aipnum4-1.bst 2010-07-25 4.21a (PWD, AO, DPC) hacked
%Control: key (0)
%Control: author (8) initials jnrlst
%Control: editor formatted (1) identically to author
%Control: production of article title (0) allowed
%Control: page (1) range
%Control: year (1) truncated
%Control: production of eprint (0) enabled
\begin{thebibliography}{36}%
\makeatletter
\providecommand \@ifxundefined [1]{%
 \@ifx{#1\undefined}
}%
\providecommand \@ifnum [1]{%
 \ifnum #1\expandafter \@firstoftwo
 \else \expandafter \@secondoftwo
 \fi
}%
\providecommand \@ifx [1]{%
 \ifx #1\expandafter \@firstoftwo
 \else \expandafter \@secondoftwo
 \fi
}%
\providecommand \natexlab [1]{#1}%
\providecommand \enquote  [1]{``#1''}%
\providecommand \bibnamefont  [1]{#1}%
\providecommand \bibfnamefont [1]{#1}%
\providecommand \citenamefont [1]{#1}%
\providecommand \href@noop [0]{\@secondoftwo}%
\providecommand \href [0]{\begingroup \@sanitize@url \@href}%
\providecommand \@href[1]{\@@startlink{#1}\@@href}%
\providecommand \@@href[1]{\endgroup#1\@@endlink}%
\providecommand \@sanitize@url [0]{\catcode `\\12\catcode `\$12\catcode
  `\&12\catcode `\#12\catcode `\^12\catcode `\_12\catcode `\%12\relax}%
\providecommand \@@startlink[1]{}%
\providecommand \@@endlink[0]{}%
\providecommand \url  [0]{\begingroup\@sanitize@url \@url }%
\providecommand \@url [1]{\endgroup\@href {#1}{\urlprefix }}%
\providecommand \urlprefix  [0]{URL }%
\providecommand \Eprint [0]{\href }%
\providecommand \doibase [0]{http://dx.doi.org/}%
\providecommand \selectlanguage [0]{\@gobble}%
\providecommand \bibinfo  [0]{\@secondoftwo}%
\providecommand \bibfield  [0]{\@secondoftwo}%
\providecommand \translation [1]{[#1]}%
\providecommand \BibitemOpen [0]{}%
\providecommand \bibitemStop [0]{}%
\providecommand \bibitemNoStop [0]{.\EOS\space}%
\providecommand \EOS [0]{\spacefactor3000\relax}%
\providecommand \BibitemShut  [1]{\csname bibitem#1\endcsname}%
\let\auto@bib@innerbib\@empty
%</preamble>
\bibitem [{\citenamefont {Pariser}(2011)}]{pariser2011filter}%
  \BibitemOpen
  \bibfield  {author} {\bibinfo {author} {\bibfnamefont {E.}~\bibnamefont
  {Pariser}},\ }\href@noop {} {\emph {\bibinfo {title} {The filter bubble: What
  the Internet is hiding from you}}}\ (\bibinfo  {publisher} {Penguin UK},\
  \bibinfo {year} {2011})\BibitemShut {NoStop}%
\bibitem [{\citenamefont {Dillahunt}, \citenamefont {Brooks},\ and\
  \citenamefont {Gulati}(2015)}]{dillahunt2015detecting}%
  \BibitemOpen
  \bibfield  {author} {\bibinfo {author} {\bibfnamefont {T.~R.}\ \bibnamefont
  {Dillahunt}}, \bibinfo {author} {\bibfnamefont {C.~A.}\ \bibnamefont
  {Brooks}}, \ and\ \bibinfo {author} {\bibfnamefont {S.}~\bibnamefont
  {Gulati}},\ }\bibfield  {title} {\enquote {\bibinfo {title} {Detecting and
  visualizing filter bubbles in google and bing},}\ }in\ \href {\doibase
  10.1145/2702613.2732850} {\emph {\bibinfo {booktitle} {Proceedings of the
  33rd Annual ACM Conference Extended Abstracts on Human Factors in Computing
  Systems}}},\ \bibinfo {series and number} {CHI EA '15}\ (\bibinfo
  {publisher} {Association for Computing Machinery},\ \bibinfo {address} {New
  York, NY, USA},\ \bibinfo {year} {2015})\ p.\ \bibinfo {pages}
  {1851–1856}\BibitemShut {NoStop}%
\bibitem [{\citenamefont {Nagulendra}\ and\ \citenamefont
  {Vassileva}(2014)}]{nagulendra2014understanding}%
  \BibitemOpen
  \bibfield  {author} {\bibinfo {author} {\bibfnamefont {S.}~\bibnamefont
  {Nagulendra}}\ and\ \bibinfo {author} {\bibfnamefont {J.}~\bibnamefont
  {Vassileva}},\ }\bibfield  {title} {\enquote {\bibinfo {title} {Understanding
  and controlling the filter bubble through interactive visualization: A user
  study},}\ }in\ \href {\doibase 10.1145/2631775.2631811} {\emph {\bibinfo
  {booktitle} {Proceedings of the 25th ACM Conference on Hypertext and Social
  Media}}},\ \bibinfo {series and number} {HT '14}\ (\bibinfo  {publisher}
  {Association for Computing Machinery},\ \bibinfo {address} {New York, NY,
  USA},\ \bibinfo {year} {2014})\ p.\ \bibinfo {pages} {107–115}\BibitemShut
  {NoStop}%
\bibitem [{\citenamefont {Nguyen}\ \emph {et~al.}(2014)\citenamefont {Nguyen},
  \citenamefont {Hui}, \citenamefont {Harper}, \citenamefont {Terveen},\ and\
  \citenamefont {Konstan}}]{nguyen2014exploring}%
  \BibitemOpen
  \bibfield  {author} {\bibinfo {author} {\bibfnamefont {T.~T.}\ \bibnamefont
  {Nguyen}}, \bibinfo {author} {\bibfnamefont {P.-M.}\ \bibnamefont {Hui}},
  \bibinfo {author} {\bibfnamefont {F.~M.}\ \bibnamefont {Harper}}, \bibinfo
  {author} {\bibfnamefont {L.}~\bibnamefont {Terveen}}, \ and\ \bibinfo
  {author} {\bibfnamefont {J.~A.}\ \bibnamefont {Konstan}},\ }\bibfield
  {title} {\enquote {\bibinfo {title} {Exploring the filter bubble: The effect
  of using recommender systems on content diversity},}\ }in\ \href {\doibase
  10.1145/2566486.2568012} {\emph {\bibinfo {booktitle} {Proceedings of the
  23rd International Conference on World Wide Web}}},\ \bibinfo {series and
  number} {WWW '14}\ (\bibinfo  {publisher} {Association for Computing
  Machinery},\ \bibinfo {address} {New York, NY, USA},\ \bibinfo {year}
  {2014})\ p.\ \bibinfo {pages} {677–686}\BibitemShut {NoStop}%
\bibitem [{\citenamefont {Bryant}(2020)}]{bryant2020youtube}%
  \BibitemOpen
  \bibfield  {author} {\bibinfo {author} {\bibfnamefont {L.~V.}\ \bibnamefont
  {Bryant}},\ }\bibfield  {title} {\enquote {\bibinfo {title} {The youtube
  algorithm and the alt-right filter bubble},}\ }\href {\doibase
  doi:10.1515/opis-2020-0007} {\bibfield  {journal} {\bibinfo  {journal} {Open
  Information Science}\ }\textbf {\bibinfo {volume} {4}},\ \bibinfo {pages}
  {85--90} (\bibinfo {year} {2020})}\BibitemShut {NoStop}%
\bibitem [{\citenamefont {O'Callaghan}\ \emph {et~al.}(2013)\citenamefont
  {O'Callaghan}, \citenamefont {Greene}, \citenamefont {Conway}, \citenamefont
  {Carthy},\ and\ \citenamefont {Cunningham}}]{o2013extreme}%
  \BibitemOpen
  \bibfield  {author} {\bibinfo {author} {\bibfnamefont {D.}~\bibnamefont
  {O'Callaghan}}, \bibinfo {author} {\bibfnamefont {D.}~\bibnamefont {Greene}},
  \bibinfo {author} {\bibfnamefont {M.}~\bibnamefont {Conway}}, \bibinfo
  {author} {\bibfnamefont {J.}~\bibnamefont {Carthy}}, \ and\ \bibinfo {author}
  {\bibfnamefont {P.}~\bibnamefont {Cunningham}},\ }\bibfield  {title}
  {\enquote {\bibinfo {title} {The extreme right filter bubble},}\ }\href@noop
  {} {\bibfield  {journal} {\bibinfo  {journal} {arXiv preprint
  arXiv:1308.6149}\ } (\bibinfo {year} {2013})}\BibitemShut {NoStop}%
\bibitem [{\citenamefont {Cinelli}\ \emph {et~al.}(2021)\citenamefont
  {Cinelli}, \citenamefont {De~Francisci~Morales}, \citenamefont {Galeazzi},
  \citenamefont {Quattrociocchi},\ and\ \citenamefont
  {Starnini}}]{Cinelli2021}%
  \BibitemOpen
  \bibfield  {author} {\bibinfo {author} {\bibfnamefont {M.}~\bibnamefont
  {Cinelli}}, \bibinfo {author} {\bibfnamefont {G.}~\bibnamefont
  {De~Francisci~Morales}}, \bibinfo {author} {\bibfnamefont {A.}~\bibnamefont
  {Galeazzi}}, \bibinfo {author} {\bibfnamefont {W.}~\bibnamefont
  {Quattrociocchi}}, \ and\ \bibinfo {author} {\bibfnamefont {M.}~\bibnamefont
  {Starnini}},\ }\bibfield  {title} {\enquote {\bibinfo {title} {The echo
  chamber effect on social media},}\ }\href {\doibase 10.1073/pnas.2023301118}
  {\bibfield  {journal} {\bibinfo  {journal} {Proceedings of the National
  Academy of Sciences}\ }\textbf {\bibinfo {volume} {118}} (\bibinfo {year}
  {2021}),\ 10.1073/pnas.2023301118},\ \Eprint
  {http://arxiv.org/abs/https://www.pnas.org/content/118/9/e2023301118.full.pdf}
  {https://www.pnas.org/content/118/9/e2023301118.full.pdf} \BibitemShut
  {NoStop}%
\bibitem [{\citenamefont {Cota}\ \emph {et~al.}(2019)\citenamefont {Cota},
  \citenamefont {Ferreira}, \citenamefont {Pastor-Satorras},\ and\
  \citenamefont {Starnini}}]{Cota2019}%
  \BibitemOpen
  \bibfield  {author} {\bibinfo {author} {\bibfnamefont {W.}~\bibnamefont
  {Cota}}, \bibinfo {author} {\bibfnamefont {S.~C.}\ \bibnamefont {Ferreira}},
  \bibinfo {author} {\bibfnamefont {R.}~\bibnamefont {Pastor-Satorras}}, \ and\
  \bibinfo {author} {\bibfnamefont {M.}~\bibnamefont {Starnini}},\ }\bibfield
  {title} {\enquote {\bibinfo {title} {Quantifying echo chamber effects in
  information spreading over political communication networks},}\ }\href
  {\doibase 10.1140/epjds/s13688-019-0213-9} {\bibfield  {journal} {\bibinfo
  {journal} {EPJ Data Science}\ }\textbf {\bibinfo {volume} {8}},\ \bibinfo
  {pages} {35} (\bibinfo {year} {2019})}\BibitemShut {NoStop}%
\bibitem [{\citenamefont {Barber\'a}\ \emph {et~al.}(2015)\citenamefont
  {Barber\'a}, \citenamefont {Jost}, \citenamefont {Nagler}, \citenamefont
  {Tucker},\ and\ \citenamefont {Bonneau}}]{barbera2015tweeting}%
  \BibitemOpen
  \bibfield  {author} {\bibinfo {author} {\bibfnamefont {P.}~\bibnamefont
  {Barber\'a}}, \bibinfo {author} {\bibfnamefont {J.~T.}\ \bibnamefont {Jost}},
  \bibinfo {author} {\bibfnamefont {J.}~\bibnamefont {Nagler}}, \bibinfo
  {author} {\bibfnamefont {J.~A.}\ \bibnamefont {Tucker}}, \ and\ \bibinfo
  {author} {\bibfnamefont {R.}~\bibnamefont {Bonneau}},\ }\bibfield  {title}
  {\enquote {\bibinfo {title} {Tweeting from left to right: Is online political
  communication more than an echo chamber?}}\ }\href {\doibase
  10.1177/0956797615594620} {\bibfield  {journal} {\bibinfo  {journal}
  {Psychological Science}\ }\textbf {\bibinfo {volume} {26}},\ \bibinfo {pages}
  {1531--1542} (\bibinfo {year} {2015})},\ \bibinfo {note} {pMID: 26297377},\
  \Eprint {http://arxiv.org/abs/https://doi.org/10.1177/0956797615594620}
  {https://doi.org/10.1177/0956797615594620} \BibitemShut {NoStop}%
\bibitem [{Pew(2017)}]{Pew2017}%
  \BibitemOpen
  \href@noop {} {\enquote {\bibinfo {title} {The partisan divide on political
  values grows even wider},}\ }\bibinfo {howpublished}
  {https://www.pewresearch.org/politics/2017/10/05/the-partisan-divide-on-political-values-grows-even-wider/}
  (\bibinfo {year} {2017}),\ \bibinfo {note} {accessed: 2021-10-04}\BibitemShut
  {NoStop}%
\bibitem [{\citenamefont {Chitra}\ and\ \citenamefont
  {Musco}(2020)}]{chitra2020analyzing}%
  \BibitemOpen
  \bibfield  {author} {\bibinfo {author} {\bibfnamefont {U.}~\bibnamefont
  {Chitra}}\ and\ \bibinfo {author} {\bibfnamefont {C.}~\bibnamefont {Musco}},\
  }\bibfield  {title} {\enquote {\bibinfo {title} {Analyzing the impact of
  filter bubbles on social network polarization},}\ }in\ \href {\doibase
  10.1145/3336191.3371825} {\emph {\bibinfo {booktitle} {Proceedings of the
  13th International Conference on Web Search and Data Mining}}},\ \bibinfo
  {series and number} {WSDM '20}\ (\bibinfo  {publisher} {Association for
  Computing Machinery},\ \bibinfo {address} {New York, NY, USA},\ \bibinfo
  {year} {2020})\ p.\ \bibinfo {pages} {115–123}\BibitemShut {NoStop}%
\bibitem [{\citenamefont {Maes}\ and\ \citenamefont
  {Bischofberger}(2015)}]{maes2015will}%
  \BibitemOpen
  \bibfield  {author} {\bibinfo {author} {\bibfnamefont {M.}~\bibnamefont
  {Maes}}\ and\ \bibinfo {author} {\bibfnamefont {L.}~\bibnamefont
  {Bischofberger}},\ }\bibfield  {title} {\enquote {\bibinfo {title} {Will the
  personalization of online social networks foster opinion polarization?}}\
  }\href@noop {} {\bibfield  {journal} {\bibinfo  {journal} {Available at SSRN
  2553436}\ } (\bibinfo {year} {2015})}\BibitemShut {NoStop}%
\bibitem [{\citenamefont {Galam}(1997)}]{galam1997rational}%
  \BibitemOpen
  \bibfield  {author} {\bibinfo {author} {\bibfnamefont {S.}~\bibnamefont
  {Galam}},\ }\bibfield  {title} {\enquote {\bibinfo {title} {Rational group
  decision making: A random field ising model at t= 0},}\ }\href@noop {}
  {\bibfield  {journal} {\bibinfo  {journal} {Physica A: Statistical Mechanics
  and its Applications}\ }\textbf {\bibinfo {volume} {238}},\ \bibinfo {pages}
  {66--80} (\bibinfo {year} {1997})}\BibitemShut {NoStop}%
\bibitem [{\citenamefont {Crokidakis}(2013)}]{crokidakis2013role}%
  \BibitemOpen
  \bibfield  {author} {\bibinfo {author} {\bibfnamefont {N.}~\bibnamefont
  {Crokidakis}},\ }\bibfield  {title} {\enquote {\bibinfo {title} {Role of
  noise and agents’ convictions on opinion spreading in a three-state
  voter-like model},}\ }\href@noop {} {\bibfield  {journal} {\bibinfo
  {journal} {Journal of Statistical Mechanics: Theory and Experiment}\ }\textbf
  {\bibinfo {volume} {2013}},\ \bibinfo {pages} {P07008} (\bibinfo {year}
  {2013})}\BibitemShut {NoStop}%
\bibitem [{\citenamefont {Ciampaglia}\ \emph {et~al.}(2018)\citenamefont
  {Ciampaglia}, \citenamefont {Nematzadeh}, \citenamefont {Menczer},\ and\
  \citenamefont {Flammini}}]{Ciampaglia2018}%
  \BibitemOpen
  \bibfield  {author} {\bibinfo {author} {\bibfnamefont {G.~L.}\ \bibnamefont
  {Ciampaglia}}, \bibinfo {author} {\bibfnamefont {A.}~\bibnamefont
  {Nematzadeh}}, \bibinfo {author} {\bibfnamefont {F.}~\bibnamefont {Menczer}},
  \ and\ \bibinfo {author} {\bibfnamefont {A.}~\bibnamefont {Flammini}},\
  }\bibfield  {title} {\enquote {\bibinfo {title} {How algorithmic popularity
  bias hinders or promotes quality},}\ }\href {\doibase
  10.1038/s41598-018-34203-2} {\bibfield  {journal} {\bibinfo  {journal}
  {Scientific Reports}\ }\textbf {\bibinfo {volume} {8}},\ \bibinfo {pages}
  {15951} (\bibinfo {year} {2018})}\BibitemShut {NoStop}%
\bibitem [{\citenamefont {Perra}\ and\ \citenamefont
  {Rocha}(2019)}]{Perra2019}%
  \BibitemOpen
  \bibfield  {author} {\bibinfo {author} {\bibfnamefont {N.}~\bibnamefont
  {Perra}}\ and\ \bibinfo {author} {\bibfnamefont {L.~E.}\ \bibnamefont
  {Rocha}},\ }\bibfield  {title} {\enquote {\bibinfo {title} {Modelling opinion
  dynamics in the age of algorithmic personalisation},}\ }\href {\doibase
  10.1038/s41598-019-43830-2} {\bibfield  {journal} {\bibinfo  {journal}
  {Scientific reports}\ }\textbf {\bibinfo {volume} {9}},\ \bibinfo {pages}
  {7261} (\bibinfo {year} {2019})}\BibitemShut {NoStop}%
\bibitem [{\citenamefont {Sirbu}\ \emph {et~al.}(2019)\citenamefont {Sirbu},
  \citenamefont {Pedreschi}, \citenamefont {Giannotti},\ and\ \citenamefont
  {Kertesz}}]{Sirbu2019}%
  \BibitemOpen
  \bibfield  {author} {\bibinfo {author} {\bibfnamefont {A.}~\bibnamefont
  {Sirbu}}, \bibinfo {author} {\bibfnamefont {D.}~\bibnamefont {Pedreschi}},
  \bibinfo {author} {\bibfnamefont {F.}~\bibnamefont {Giannotti}}, \ and\
  \bibinfo {author} {\bibfnamefont {J.}~\bibnamefont {Kertesz}},\ }\bibfield
  {title} {\enquote {\bibinfo {title} {Algorithmic bias amplifies opinion
  fragmentation and polarization: A bounded confidence model},}\ }\href
  {\doibase 10.1371/journal.pone.0213246} {\bibfield  {journal} {\bibinfo
  {journal} {PLOS ONE}\ }\textbf {\bibinfo {volume} {14}},\ \bibinfo {pages}
  {1--20} (\bibinfo {year} {2019})}\BibitemShut {NoStop}%
\bibitem [{\citenamefont {Freitas}, \citenamefont {Vieira},\ and\ \citenamefont
  {Anteneodo}(2020)}]{freitas2020imperfect}%
  \BibitemOpen
  \bibfield  {author} {\bibinfo {author} {\bibfnamefont {F.}~\bibnamefont
  {Freitas}}, \bibinfo {author} {\bibfnamefont {A.~R.}\ \bibnamefont {Vieira}},
  \ and\ \bibinfo {author} {\bibfnamefont {C.}~\bibnamefont {Anteneodo}},\
  }\bibfield  {title} {\enquote {\bibinfo {title} {Imperfect bifurcations in
  opinion dynamics under external fields},}\ }\href@noop {} {\bibfield
  {journal} {\bibinfo  {journal} {Journal of Statistical Mechanics: Theory and
  Experiment}\ }\textbf {\bibinfo {volume} {2020}},\ \bibinfo {pages} {024002}
  (\bibinfo {year} {2020})}\BibitemShut {NoStop}%
\bibitem [{\citenamefont {Baumann}\ \emph {et~al.}(2020)\citenamefont
  {Baumann}, \citenamefont {Lorenz-Spreen}, \citenamefont {Sokolov},\ and\
  \citenamefont {Starnini}}]{Baumann2020}%
  \BibitemOpen
  \bibfield  {author} {\bibinfo {author} {\bibfnamefont {F.}~\bibnamefont
  {Baumann}}, \bibinfo {author} {\bibfnamefont {P.}~\bibnamefont
  {Lorenz-Spreen}}, \bibinfo {author} {\bibfnamefont {I.~M.}\ \bibnamefont
  {Sokolov}}, \ and\ \bibinfo {author} {\bibfnamefont {M.}~\bibnamefont
  {Starnini}},\ }\bibfield  {title} {\enquote {\bibinfo {title} {Modeling echo
  chambers and polarization dynamics in social networks},}\ }\href {\doibase
  10.1103/PhysRevLett.124.048301} {\bibfield  {journal} {\bibinfo  {journal}
  {Phys. Rev. Lett.}\ }\textbf {\bibinfo {volume} {124}},\ \bibinfo {pages}
  {048301} (\bibinfo {year} {2020})}\BibitemShut {NoStop}%
\bibitem [{\citenamefont {Peralta}\ \emph {et~al.}(2021)\citenamefont
  {Peralta}, \citenamefont {Neri}, \citenamefont {Kert\'esz},\ and\
  \citenamefont {I\~niguez}}]{Peralta2021}%
  \BibitemOpen
  \bibfield  {author} {\bibinfo {author} {\bibfnamefont {A.~F.}\ \bibnamefont
  {Peralta}}, \bibinfo {author} {\bibfnamefont {M.}~\bibnamefont {Neri}},
  \bibinfo {author} {\bibfnamefont {J.}~\bibnamefont {Kert\'esz}}, \ and\
  \bibinfo {author} {\bibfnamefont {G.}~\bibnamefont {I\~niguez}},\ }\bibfield
  {title} {\enquote {\bibinfo {title} {Effect of algorithmic bias and network
  structure on coexistence, consensus, and polarization of opinions},}\ }\href
  {\doibase 10.1103/PhysRevE.104.044312} {\bibfield  {journal} {\bibinfo
  {journal} {Phys. Rev. E}\ }\textbf {\bibinfo {volume} {104}},\ \bibinfo
  {pages} {044312} (\bibinfo {year} {2021})}\BibitemShut {NoStop}%
\bibitem [{\citenamefont {Baron}(2021)}]{Baron2021}%
  \BibitemOpen
  \bibfield  {author} {\bibinfo {author} {\bibfnamefont {J.~W.}\ \bibnamefont
  {Baron}},\ }\bibfield  {title} {\enquote {\bibinfo {title} {Consensus,
  polarization, and coexistence in a continuous opinion dynamics model with
  quenched disorder},}\ }\href {\doibase 10.1103/PhysRevE.104.044309}
  {\bibfield  {journal} {\bibinfo  {journal} {Phys. Rev. E}\ }\textbf {\bibinfo
  {volume} {104}},\ \bibinfo {pages} {044309} (\bibinfo {year}
  {2021})}\BibitemShut {NoStop}%
\bibitem [{\citenamefont {Castellano}, \citenamefont {Fortunato},\ and\
  \citenamefont {Loreto}(2009)}]{Castellano2009}%
  \BibitemOpen
  \bibfield  {author} {\bibinfo {author} {\bibfnamefont {C.}~\bibnamefont
  {Castellano}}, \bibinfo {author} {\bibfnamefont {S.}~\bibnamefont
  {Fortunato}}, \ and\ \bibinfo {author} {\bibfnamefont {V.}~\bibnamefont
  {Loreto}},\ }\bibfield  {title} {\enquote {\bibinfo {title} {Statistical
  physics of social dynamics},}\ }\href {\doibase 10.1103/RevModPhys.81.591}
  {\bibfield  {journal} {\bibinfo  {journal} {Rev. Mod. Phys.}\ }\textbf
  {\bibinfo {volume} {81}},\ \bibinfo {pages} {591--646} (\bibinfo {year}
  {2009})}\BibitemShut {NoStop}%
\bibitem [{\citenamefont {Sen}\ and\ \citenamefont
  {Chakrabarti}(2014)}]{Sen2014}%
  \BibitemOpen
  \bibfield  {author} {\bibinfo {author} {\bibfnamefont {P.}~\bibnamefont
  {Sen}}\ and\ \bibinfo {author} {\bibfnamefont {B.~K.}\ \bibnamefont
  {Chakrabarti}},\ }\href@noop {} {\emph {\bibinfo {title} {Sociophysics: an
  introduction}}}\ (\bibinfo  {publisher} {Oxford University Press},\ \bibinfo
  {year} {2014})\BibitemShut {NoStop}%
\bibitem [{\citenamefont {Clifford}\ and\ \citenamefont
  {Sudbury}(1973)}]{Clifford1973}%
  \BibitemOpen
  \bibfield  {author} {\bibinfo {author} {\bibfnamefont {P.}~\bibnamefont
  {Clifford}}\ and\ \bibinfo {author} {\bibfnamefont {A.}~\bibnamefont
  {Sudbury}},\ }\bibfield  {title} {\enquote {\bibinfo {title} {A model for
  spatial conflict},}\ }\href {\doibase 10.1093/biomet/60.3.581} {\bibfield
  {journal} {\bibinfo  {journal} {Biometrika}\ }\textbf {\bibinfo {volume}
  {60}},\ \bibinfo {pages} {581--588} (\bibinfo {year} {1973})}\BibitemShut
  {NoStop}%
\bibitem [{\citenamefont {Frachebourg}\ and\ \citenamefont
  {Krapivsky}(1996)}]{Frachebourg1996}%
  \BibitemOpen
  \bibfield  {author} {\bibinfo {author} {\bibfnamefont {L.}~\bibnamefont
  {Frachebourg}}\ and\ \bibinfo {author} {\bibfnamefont {P.~L.}\ \bibnamefont
  {Krapivsky}},\ }\bibfield  {title} {\enquote {\bibinfo {title} {Exact results
  for kinetics of catalytic reactions},}\ }\href {\doibase
  10.1103/PhysRevE.53.R3009} {\bibfield  {journal} {\bibinfo  {journal} {Phys.
  Rev. E}\ }\textbf {\bibinfo {volume} {53}},\ \bibinfo {pages} {R3009--R3012}
  (\bibinfo {year} {1996})}\BibitemShut {NoStop}%
\bibitem [{\citenamefont {Krapivsky}, \citenamefont {Redner},\ and\
  \citenamefont {Ben-Naim}(2010)}]{Rednerbook}%
  \BibitemOpen
  \bibfield  {author} {\bibinfo {author} {\bibfnamefont {P.~L.}\ \bibnamefont
  {Krapivsky}}, \bibinfo {author} {\bibfnamefont {S.}~\bibnamefont {Redner}}, \
  and\ \bibinfo {author} {\bibfnamefont {E.}~\bibnamefont {Ben-Naim}},\
  }\href@noop {} {\emph {\bibinfo {title} {A kinetic view of statistical
  physics}}}\ (\bibinfo  {publisher} {Cambridge University Press},\ \bibinfo
  {year} {2010})\BibitemShut {NoStop}%
\bibitem [{\citenamefont {Starnini}, \citenamefont {Baronchelli},\ and\
  \citenamefont {Pastor-Satorras}(2012)}]{Starnini2012}%
  \BibitemOpen
  \bibfield  {author} {\bibinfo {author} {\bibfnamefont {M.}~\bibnamefont
  {Starnini}}, \bibinfo {author} {\bibfnamefont {A.}~\bibnamefont
  {Baronchelli}}, \ and\ \bibinfo {author} {\bibfnamefont {R.}~\bibnamefont
  {Pastor-Satorras}},\ }\bibfield  {title} {\enquote {\bibinfo {title}
  {Ordering dynamics of the multi-state voter model},}\ }\href {\doibase
  10.1088/1742-5468/2012/10/p10027} {\bibfield  {journal} {\bibinfo  {journal}
  {Journal of Statistical Mechanics: Theory and Experiment}\ }\textbf {\bibinfo
  {volume} {2012}},\ \bibinfo {pages} {P10027} (\bibinfo {year}
  {2012})}\BibitemShut {NoStop}%
\bibitem [{\citenamefont {Pickering}\ and\ \citenamefont
  {Lim}(2016)}]{Pickering2016}%
  \BibitemOpen
  \bibfield  {author} {\bibinfo {author} {\bibfnamefont {W.}~\bibnamefont
  {Pickering}}\ and\ \bibinfo {author} {\bibfnamefont {C.}~\bibnamefont
  {Lim}},\ }\bibfield  {title} {\enquote {\bibinfo {title} {Solution of the
  multistate voter model and application to strong neutrals in the naming
  game},}\ }\href {\doibase 10.1103/PhysRevE.93.032318} {\bibfield  {journal}
  {\bibinfo  {journal} {Phys. Rev. E}\ }\textbf {\bibinfo {volume} {93}},\
  \bibinfo {pages} {032318} (\bibinfo {year} {2016})}\BibitemShut {NoStop}%
\bibitem [{\citenamefont {Peralta}, \citenamefont {Khalil},\ and\ \citenamefont
  {Toral}(2020)}]{Peralta2020}%
  \BibitemOpen
  \bibfield  {author} {\bibinfo {author} {\bibfnamefont {A.~F.}\ \bibnamefont
  {Peralta}}, \bibinfo {author} {\bibfnamefont {N.}~\bibnamefont {Khalil}}, \
  and\ \bibinfo {author} {\bibfnamefont {R.}~\bibnamefont {Toral}},\ }\bibfield
   {title} {\enquote {\bibinfo {title} {Ordering dynamics in the voter model
  with aging},}\ }\href {\doibase https://doi.org/10.1016/j.physa.2019.122475}
  {\bibfield  {journal} {\bibinfo  {journal} {Physica A: Statistical Mechanics
  and its Applications}\ }\textbf {\bibinfo {volume} {552}},\ \bibinfo {pages}
  {122475} (\bibinfo {year} {2020})},\ \bibinfo {note} {tributes of
  Non-equilibrium Statistical Physics}\BibitemShut {NoStop}%
\bibitem [{\citenamefont {De~Marzo}, \citenamefont {Zaccaria},\ and\
  \citenamefont {Castellano}(2020)}]{DeMarzo2020}%
  \BibitemOpen
  \bibfield  {author} {\bibinfo {author} {\bibfnamefont {G.}~\bibnamefont
  {De~Marzo}}, \bibinfo {author} {\bibfnamefont {A.}~\bibnamefont {Zaccaria}},
  \ and\ \bibinfo {author} {\bibfnamefont {C.}~\bibnamefont {Castellano}},\
  }\bibfield  {title} {\enquote {\bibinfo {title} {Emergence of polarization in
  a voter model with personalized information},}\ }\href {\doibase
  10.1103/PhysRevResearch.2.043117} {\bibfield  {journal} {\bibinfo  {journal}
  {Phys. Rev. Research}\ }\textbf {\bibinfo {volume} {2}},\ \bibinfo {pages}
  {043117} (\bibinfo {year} {2020})}\BibitemShut {NoStop}%
\bibitem [{\citenamefont {Herrer\'{\i}as-Azcu\'e}\ and\ \citenamefont
  {Galla}(2019)}]{herrerias2019consensus}%
  \BibitemOpen
  \bibfield  {author} {\bibinfo {author} {\bibfnamefont {F.}~\bibnamefont
  {Herrer\'{\i}as-Azcu\'e}}\ and\ \bibinfo {author} {\bibfnamefont
  {T.}~\bibnamefont {Galla}},\ }\bibfield  {title} {\enquote {\bibinfo {title}
  {Consensus and diversity in multistate noisy voter models},}\ }\href
  {\doibase 10.1103/PhysRevE.100.022304} {\bibfield  {journal} {\bibinfo
  {journal} {Phys. Rev. E}\ }\textbf {\bibinfo {volume} {100}},\ \bibinfo
  {pages} {022304} (\bibinfo {year} {2019})}\BibitemShut {NoStop}%
\bibitem [{Note1()}]{Note1}%
  \BibitemOpen
  \bibinfo {note} {Note that these transition probabilities do not depend on
  the state $\sigma _i$ of the agent}\BibitemShut {NoStop}%
\bibitem [{\citenamefont {Schafer}\ \emph {et~al.}(2007)\citenamefont
  {Schafer}, \citenamefont {Frankowski}, \citenamefont {Herlocker},\ and\
  \citenamefont {Sen}}]{schafer2007collaborative}%
  \BibitemOpen
  \bibfield  {author} {\bibinfo {author} {\bibfnamefont {J.~B.}\ \bibnamefont
  {Schafer}}, \bibinfo {author} {\bibfnamefont {D.}~\bibnamefont {Frankowski}},
  \bibinfo {author} {\bibfnamefont {J.}~\bibnamefont {Herlocker}}, \ and\
  \bibinfo {author} {\bibfnamefont {S.}~\bibnamefont {Sen}},\ }\bibfield
  {title} {\enquote {\bibinfo {title} {Collaborative filtering recommender
  systems},}\ }in\ \href@noop {} {\emph {\bibinfo {booktitle} {The adaptive
  web}}}\ (\bibinfo  {publisher} {Springer},\ \bibinfo {year} {2007})\ pp.\
  \bibinfo {pages} {291--324}\BibitemShut {NoStop}%
\bibitem [{\citenamefont {Linden}, \citenamefont {Smith},\ and\ \citenamefont
  {York}(2003)}]{linden2003amazon}%
  \BibitemOpen
  \bibfield  {author} {\bibinfo {author} {\bibfnamefont {G.}~\bibnamefont
  {Linden}}, \bibinfo {author} {\bibfnamefont {B.}~\bibnamefont {Smith}}, \
  and\ \bibinfo {author} {\bibfnamefont {J.}~\bibnamefont {York}},\ }\bibfield
  {title} {\enquote {\bibinfo {title} {Amazon.com recommendations: item-to-item
  collaborative filtering},}\ }\href {\doibase 10.1109/MIC.2003.1167344}
  {\bibfield  {journal} {\bibinfo  {journal} {IEEE Internet Computing}\
  }\textbf {\bibinfo {volume} {7}},\ \bibinfo {pages} {76--80} (\bibinfo {year}
  {2003})}\BibitemShut {NoStop}%
\bibitem [{\citenamefont {Smith}\ and\ \citenamefont
  {Linden}(2017)}]{smith2017two}%
  \BibitemOpen
  \bibfield  {author} {\bibinfo {author} {\bibfnamefont {B.}~\bibnamefont
  {Smith}}\ and\ \bibinfo {author} {\bibfnamefont {G.}~\bibnamefont {Linden}},\
  }\bibfield  {title} {\enquote {\bibinfo {title} {Two decades of recommender
  systems at amazon.com},}\ }\href {\doibase 10.1109/MIC.2017.72} {\bibfield
  {journal} {\bibinfo  {journal} {IEEE Internet Computing}\ }\textbf {\bibinfo
  {volume} {21}},\ \bibinfo {pages} {12--18} (\bibinfo {year}
  {2017})}\BibitemShut {NoStop}%
\bibitem [{Note2()}]{Note2}%
  \BibitemOpen
  \bibinfo {note} {Only two because the scaling $\lambda <N^{-1/2}$ corresponds
  to negative times.}\BibitemShut {Stop}%
\end{thebibliography}%

\appendix
%\section{The case $c=1$}
\section{Moments of the distribution of $n_i^{(k)}$}\label{apx:Master equation, transition rates and moments}
\begin{figure*}[t]
	\centering
	\begin{subfigure}{\textwidth}
		\centering
		\includegraphics[width=.81\textwidth]{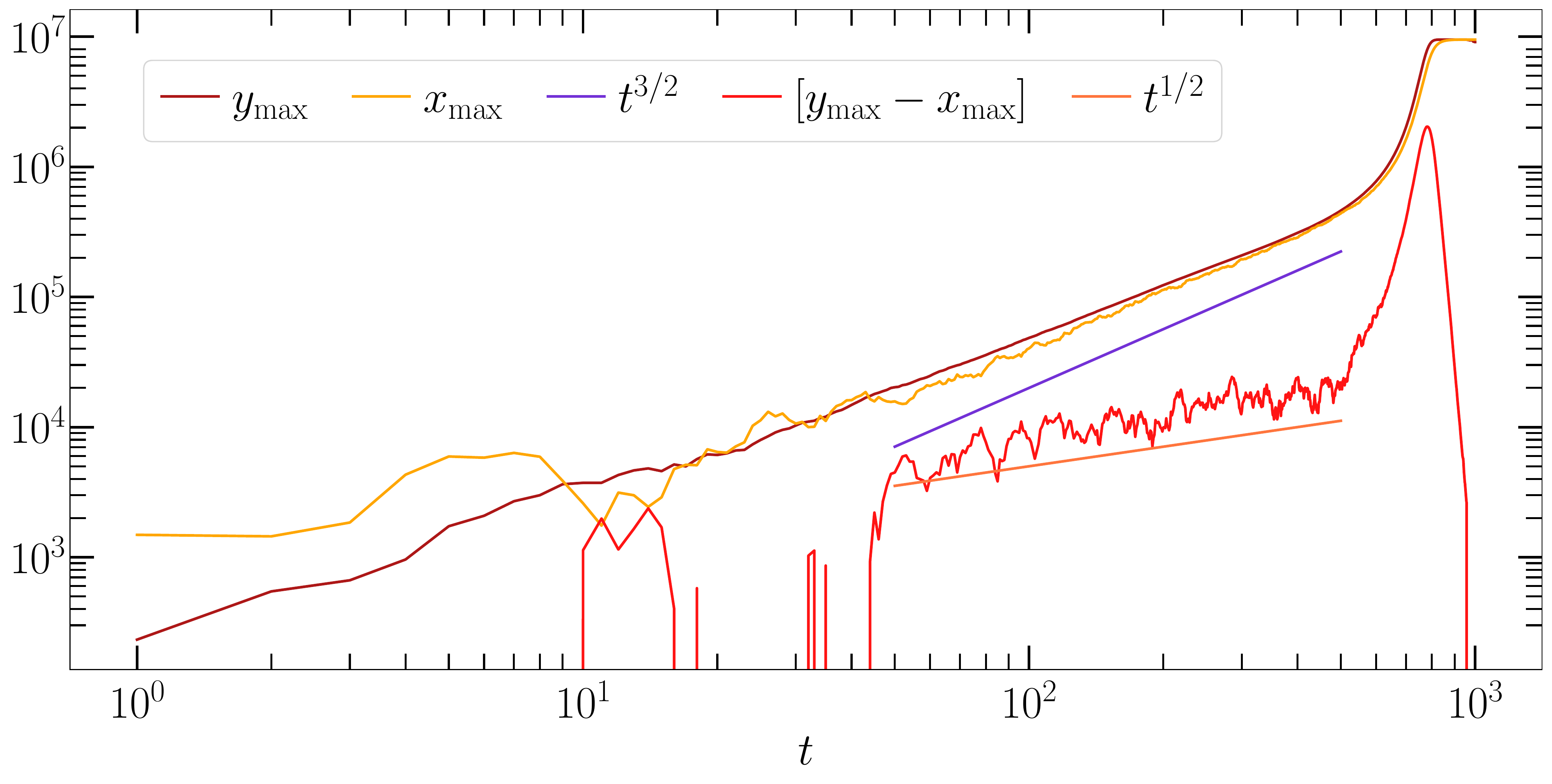}
		\caption{}
	\end{subfigure}\hfill%
	\begin{subfigure}{.45\textwidth}
		\centering
		\includegraphics[width=.9\textwidth]{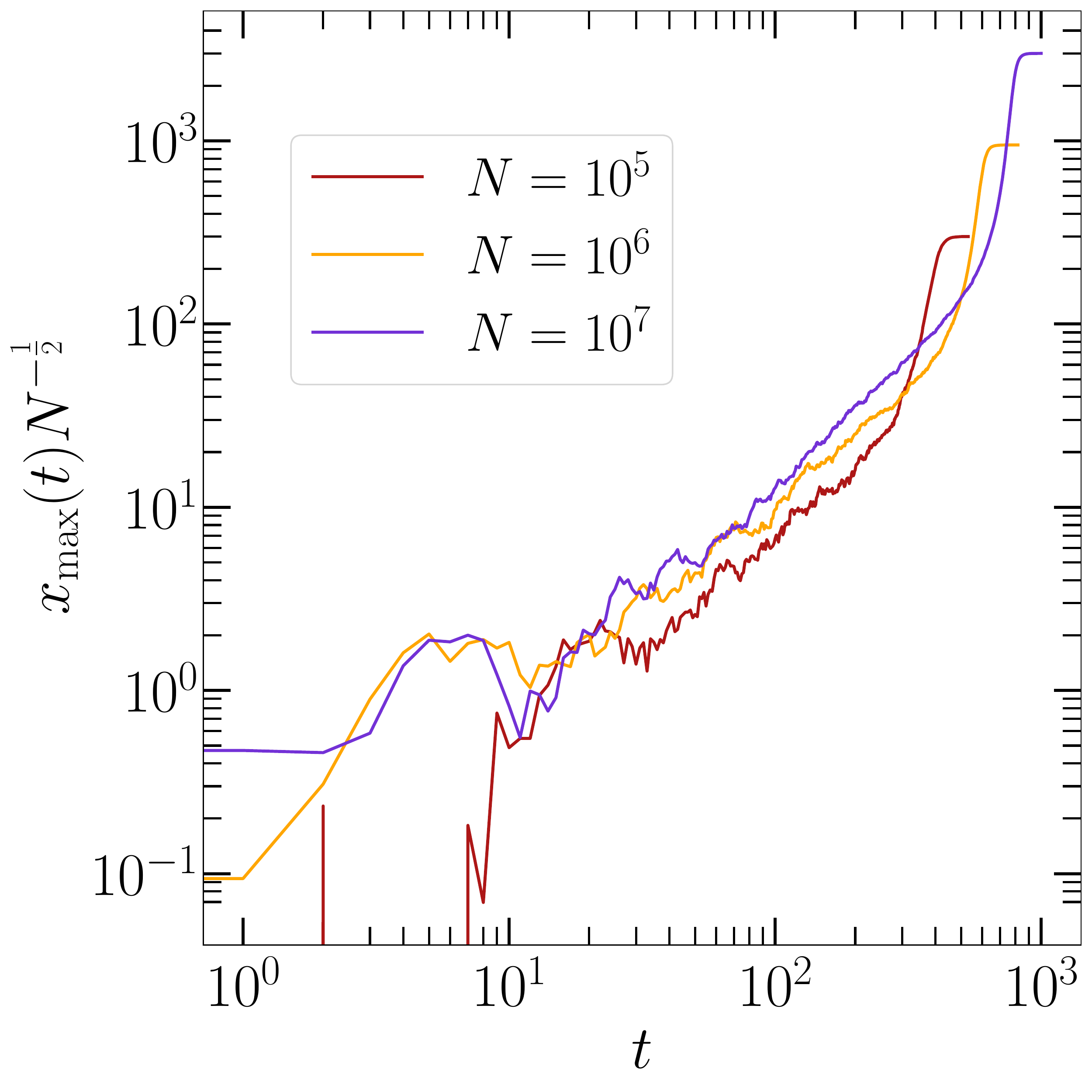}
		\caption{}
	\end{subfigure}
	\begin{subfigure}{.45\textwidth}
		\centering
		\includegraphics[width=.9\textwidth]{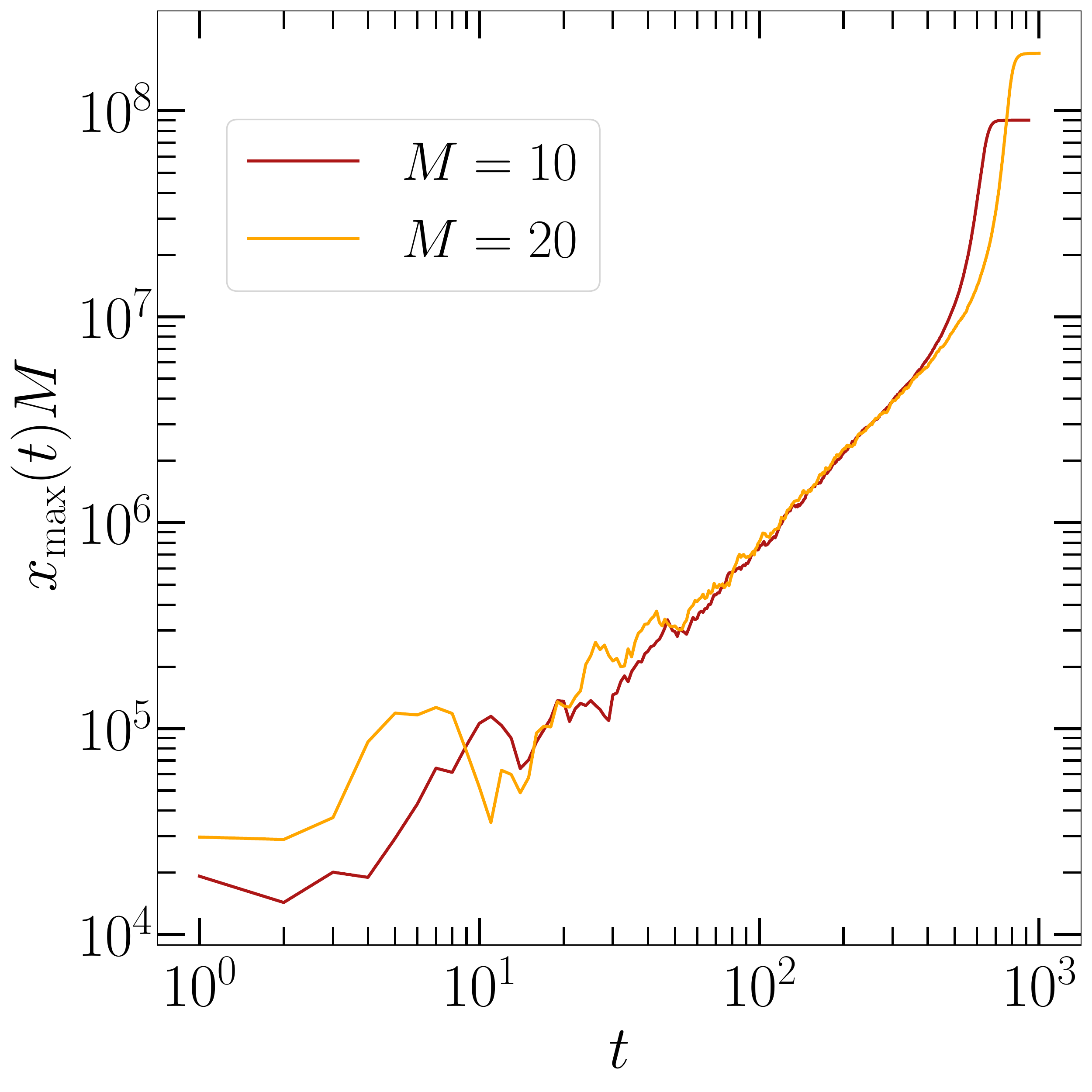}
		\caption{}
	\end{subfigure}\hfill%
	\caption{(a) Temporal evolution of $x_{\max}(t)$, $y_{\max}(t)$ and their difference for $N=10^7$, $M=20$, $\lambda \approx \lambda_c$.
          (b) Collapse plot to check the scaling~\eqref{eq:x_max(t)} with
          respect to $N$; $M=20$ and $\lambda \approx \lambda_c$.
          (c) Collapse plot to check the scaling~\eqref{eq:x_max(t)} with
        respect to $M$; $N=10^7$ and $\lambda \approx \lambda_c$.}
        \label{fig:scaling_x_max}
\end{figure*}
Let us consider the distribution $Q\ton*{n_i^{(k)},t}$, where
$n_i^{(k)}$ is defined as the number of times agent $i$ has chosen
or confirmed opinion $k$ in the past.
The evolution of this distribution is given by
\begin{align*}
  Q\ton*{n_i^{(k)},t+\delta t}&=R\ton*{n_i^{(k)}-1}Q\ton*{n_i^{(k)}-1,t}+\\
  &+\qua*{1-R\ton*{n_i^{(k)}}}Q\ton*{n_i^{(k)},t}.
\end{align*}
Expanding the left hand side for small $\delta t=1/N$ we obtain 
%\begin{align*}
%  Q\ton*{n_i^{(k)},t}+\frac{dQ\ton*{n_i^{(k)},t}}{dt}\delta t&=R\ton*{n_i^{(k)}-1}Q\ton*{n_i^{(k)}-1,t}+\\
%  &+\qua*{1-R\ton*{n_i^{(k)}}}Q\ton*{n_i^{(k)},t},
%\end{align*}
which yields Eq.~\eqref{eq:master_equation_n_general}
\begin{align}
  \frac{dQ\ton*{n_i^{(k)}}}{dt}&=NR\ton*{n_i^{(k)}-1}Q\ton*{n_i^{(k)}-1}- \nonumber \\
  &-NR\ton*{n_i^{(k)}}Q\ton*{n_i^{(k)}}.
  \label{eq:master_equation_n_general_appendix}
\end{align}

Now we consider the two components $Q_1\ton*{n_i^{(k)},t}$ and
$Q_2\ton*{n_i^{(k)},t}$ introduced in \ref{subsec:polarization_time}
and we compute their mean value and variance. We recall that $Q_2$
corresponds to agents whose PI are polarized along $k$, while $Q_1$ to
the remaining agents, and that both components evolve according to the
same master equation Eq.~\eqref{eq:master_equation_n_general}, but
with different transition rates.  The general equation for the drift of the distribution (also called average drift in the following)
is
\[
\nu=\frac{d\mean*{n_i^{(k)}}}{dt}=\sum_{n_i^{k}}n_i^{(k)}\frac{dQ\ton*{n_i^{(k)}}}{dt}
\]
and using Eq.~\eqref{eq:master_equation_n_general_appendix} we obtain 
\begin{align}
  \nu=\sum_{n_i^{k}}&n_i^{(k)}\left[NR\ton*{n_i^{(k)}-1}Q\ton*{n_i^{(k)}-1}\right.-\nonumber\\
    &-\left.NR\ton*{n_i^{(k)}}Q\ton*{n_i^{(k)}}\right]=\nonumber\\
  =&\sum_{n_i^{k}}NR\ton*{n_i^{(k)}}Q\ton*{n_i^{(k)}}
  \label{eq:drift_general}
\end{align} 
Analogously the evolution of the second moment is 
\[
\frac{d\mean*{\ton*{n_i^{(k)}}^2}}{dt}=\sum_{n_i^{k}}\ton*{n_i^{(k)}}^2\frac{dQ\ton*{n_i^{(k)}}}{dt},
\]
that using Eq.~\eqref{eq:master_equation_n_general_appendix} becomes
\begin{align}
  \frac{d\mean*{\ton*{n_i^{(k)}}^2}}{dt}=\sum_{n_i^{k}}&\ton*{n_i^{(k)}}^2\left[NR\ton*{n_i^{(k)}-1}Q\ton*{n_i^{(k)}-1}\right.-\nonumber\\
    &-\left.NR\ton*{n_i^{(k)}}Q\ton*{n_i^{(k)}}\right]=\nonumber\\
  =\sum_{n_i^{k}}&\ton*{n_i^{(k)}+1}^2NR\ton*{n_i^{(k)}}Q\ton*{n_i^{(k)}}-\nonumber\\
  &-\ton*{n_i^{(k)}}^2NR\ton*{n_i^{(k)}}Q\ton*{n_i^{(k)}}=\nonumber\\
  =&\nu+2\sum_{n_i^{k}}n_i^{(k)}NR\ton*{n_i^{(k)}}Q\ton*{n_i^{(k)}},
  \label{eq:second_moment_general}
\end{align} 
where we used Eq.~\eqref{eq:drift_general}.
Let us firstly consider agents polarized along $k$ and so $Q_2$, the transition rate is 
\begin{equation*}
  NR_2\ton*{n_i^{(k)}}=(1-\lambda)\frac{1}{M_s(t)}+\lambda.
  \label{eq:transition_2}
\end{equation*}
and putting this expression into Eqs.~\eqref{eq:drift_general} and \eqref{eq:second_moment_general} we get
\begin{equation}
  \begin{dcases}
    \nu_2=(1-\lambda)\frac{1}{M_s(t)}+\lambda\\
    \frac{d\mean*{\ton*{n_i^{(k)}}^2}_2}{dt}=\qua*{(1-\lambda)\frac{1}{M_s(t)}+\lambda}\ton*{1+2n_2},
  \end{dcases}
  \label{eq:nu_mean_n^2_2}
\end{equation}
where we defined $n_2=\mean*{n_i^{(k)}}_2$. Analogously for the component $Q_1$
it holds 
\[
NR_1\ton*{n_i^{(k)}}=(1-\lambda)\frac{1}{M_s(t)}
\]
and so
\begin{equation}
  \begin{dcases}
    \nu_1=(1-\lambda)\frac{1}{M_s(t)}\\
    \frac{d\mean*{\ton*{n_i^{(k)}}^2}_1}{dt}=(1-\lambda)\frac{1}{M_s(t)}\ton*{1+2n_2}.
  \end{dcases}
  \label{eq:nu_mean_n^2_1}
\end{equation}
Finally, considering that 
\begin{align*}
  &n=\int_0^t \nu \cdot dt'\\
  &\sigma_n^2=\mean*{\ton*{n_i^{(k)}}^2}-\mean{n_i^{(k)}}^2\\
  &\frac{d\sigma^2}{dt}=\frac{d\mean*{\ton*{n_i^{(k)}}^2}}{dt}-2\mean*{n_i^{(k)}}\frac{d\mean*{n_i^{(k)}}}{dt}
\end{align*}
we can write the expressions for the mean value and variance of both components
\begin{equation}
  \begin{dcases}
    n_2(t)=\sigma_2^2(t)=\int_0^tdt'\ton*{\frac{1-\lambda}{M_s(t')}+\lambda}\\
    n_1(t)=\sigma_1^2(t)=\int_0^tdt'\frac{1-\lambda}{M_s(t')}.\\
  \end{dcases}
\end{equation}
These are Eqs.~\eqref{eq:mean_variance_2} and \eqref{eq:mean_variance_1}.

\section{Scaling of $x_{\max}(t)$}
\begin{figure}[t]
  \includegraphics[width=\linewidth]{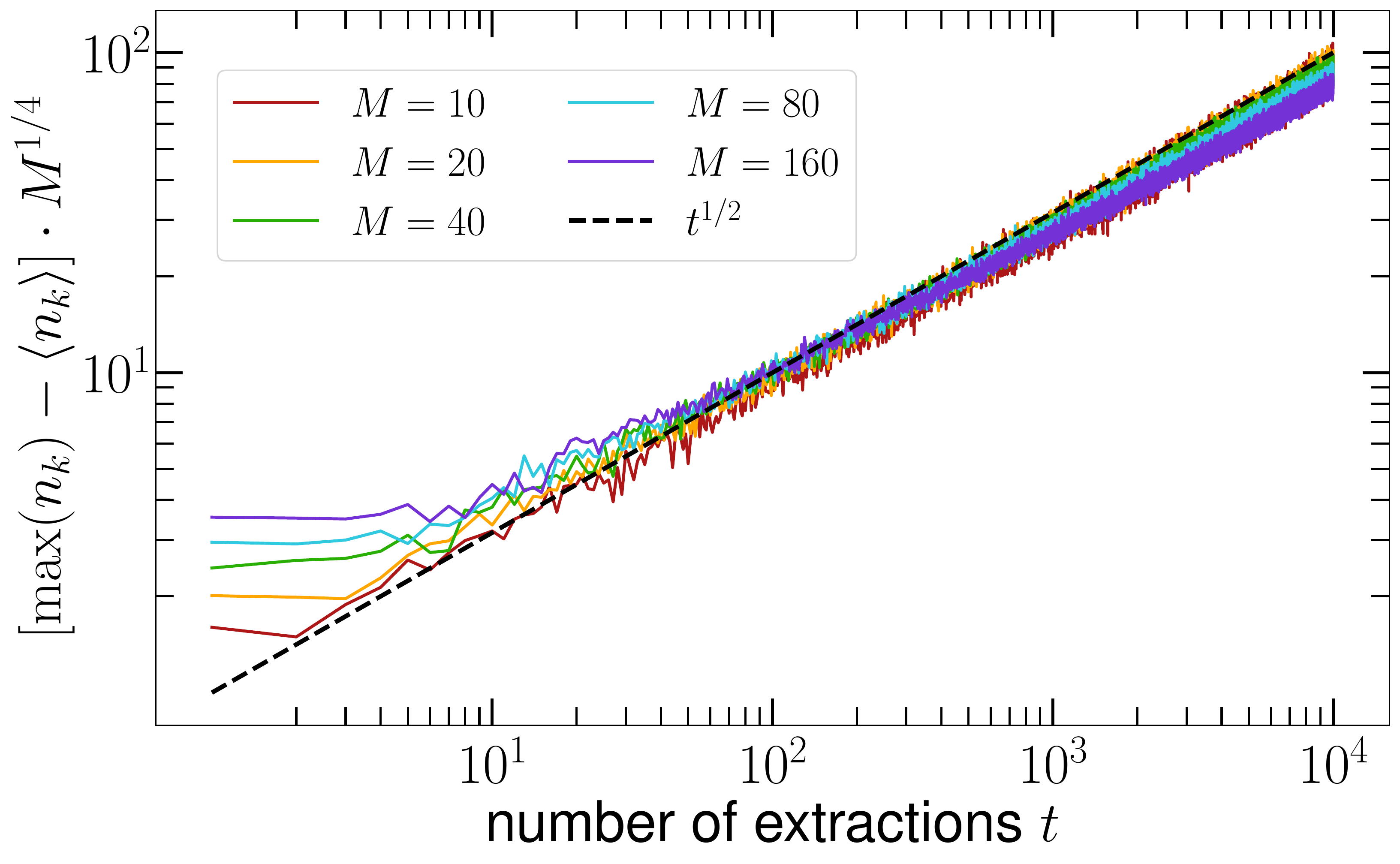}
  \caption{Growth of the maximum of a multinomial distribution with
    $M$ equally probable classes as a function of the number of
    extractions $t$ for different values of $M$. We denote by $n_k$
    the number of counts in the $k$th classe. We plotted the quantity
    $\left\{\max\left[n_k\right]-\mean{n_k}\right\}M^{1/4}$ to highlight
    the scaling with respect to $M$ and we averaged over $25$
    different realizations of the process.}
  \label{fig:multinomial}
\end{figure}
\label{apx:Scaling of x_max}
As discussed in the main text, near the transition point the quantity
$x_{\max}(t)=N_{\max}(t)-N/M$ satisfies Eq.~\eqref{eq:x_max(t)}, that is
\[
x_{\max}(t)=\gamma\frac{N^{1/2}}{M}t^{3/2}.
\]
This is shown in Fig.~\ref{fig:scaling_x_max}, where we test both the
evolution in time and the scaling with respect to $M$ and $N$.  The
latter can be also explained by considering the dynamics during the
first steps. Initially all $n_i^{(k)}$ are null and all opinions are
equally common, so each agent, when updated, chooses an opinion with
uniform probability. This regime lasts up to $t<\bar{t}$, where
$\bar{t}$ is defined as the time when the $n_i^{(k)}$ becomes
statistically different from zero and so the PI stops to be completely
random. This implies that for fixed $i$ and up to $\bar{t}$ the $M$ variables
$n_i^{(k)}$ are distributed according to a multinomial distribution
with uniform probability $1/M$. The multinomial regime ends when the
maximum $\max_k\left[n_i^{(k)}\right]$ exceeds the average
$\mean{n_i^{(k)}}_k$ by one, since when this occurs the external
information stops to suggest random opinions. The mean value after $t$
extractions satisfies $\mean{n_i^{(k)}}_k=t/M$, while by
drawing from a multinomial we numerically determined that
$\max_k\left[n_i^{(k)}\right]-\mean{n_i^{(k)}}_k\approx
t^{1/2}M^{-1/4}$, see Fig.~\ref{fig:multinomial}.
We can then obtain the time $\bar{t}$ as
\begin{equation}
  \bar{t}^{1/2}M^{-1/4}=1 \ \to \ \bar{t}=M^{1/2}.
  \label{eq:bar_t}
\end{equation}

Now we consider the evolution of the number of agents $N_k$ in opinion
$k$ in the simple multistate voter model. The transition rates are
\begin{equation}
  \begin{dcases}
    W\ton*{N_k\to N_k\pm1}=\frac{1}{\delta t}\frac{N-N_k}{N}\frac{N_k}{N}\\
    W\ton*{N_k\pm1\to N_k}=\frac{1}{\delta t}\frac{N-\ton*{N_k\pm1}}{N}\frac{N_k\pm1}{N},
  \end{dcases}
  \label{eq:transition_rates_N_k}
\end{equation}
where $\delta t=1/N$. The master equation for $N_k$ is thus 
\begin{align*}
  P(N_k,t+&\delta t)=P(N_k+1,t)W\ton*{N_k+1\to N_k}+\\
  &+P(N_k-1,t)W\ton*{N_k-1\to N_k}+\\
  &+\qua*{1-W\ton*{N_k\to N_k-1}-W\ton*{N_k\to N_k+1}}P\ton*{N_k, t}.
\end{align*}
Expanding for $\delta t$ small we get
\begin{align}
  \frac{\partial P(N_k, t)}{\partial t} &\cdot \delta t
  = W(N_k+1 \to N_k) P(N_k+1,t)\nonumber\\
  +& W(N_k-1 \to N_k) P(N_k-1,t)+\nonumber\\
  -& [W(N_k+1 \to N_k) + W(N_k-1 \to N_k)] P(N_k,t).
  \label{eq:master_equation_N_k}
\end{align}

Starting from this master equation we can compute the average drift and the variance of $N_k$.
The average drift satisfies 
\begin{align}
\nu_k=&\frac{d\mean{N_k}}{dt}=\sum_{N_k}N_k\frac{\partial P(N_k, t)}{\partial t} \nonumber \\
=& \frac{1}{\delta t}\sum_{N_k} [W(N_k \to N_k+1) - W(N_k \to N_k-1)] P(N_k,t)=\nonumber\\
=& \sum_{N_k}  \frac{dN_k}{dt}P(N_k,t)=\sum_{N_k} \nu_k(N_k)P(N_k,t),
\label{drift}
\end{align}
where we also introduced the drift $\nu_k(N_k)=[W(N_k \to N_k+1) -
  W(N_k \to N_k-1)]/\delta t$. Note that while the average drift
$\nu_k$ determines how the mean value of the distribution $P(N_k)$
moves in time, the drift $\nu_k(N_k)$ allows to compute (neglecting
diffusion) how a specific value of $N_k$ evolves.  Using
Eqs.~\eqref{eq:transition_rates_N_k} we get
\begin{equation}
  \nu_k=0 \ \to \ \mean*{N_k(t)}=N_k(0)=\frac{N}{M}.
  \label{eq:drift_evolution_N_k}
\end{equation}
Analogously the evolution of the variance is 
\[
\frac{d\sigma_k^2}{dt}=\frac{d\mean*{N_k^2}}{dt}-2\mean*{N_k}\frac{d\mean*{N_k}}{dt}=\frac{d\mean*{N_k^2}}{dt},
\]
where the last equality follows from
Eq.~\eqref{eq:drift_evolution_N_k}. Exploiting
Eq.~\eqref{eq:master_equation_N_k} we then obtain
\[
\frac{d\mean*{N_k^2}}{dt} = \sum_{N_k}N_k^2\frac{\partial P(N_k, t)}{\partial t}=2\ton*{\mean*{N_k}-\frac{1}{N}\mean*{N_k^2}},
\]
whose solution is
\[
\mean*{N_k^2}(t)=\ton*{\frac{N^2}{M^2}-\frac{N^2}{M}}\me^{-2\frac{t}{N}}+\frac{N^2}{M}.
\]
The variance is thus 
\begin{equation}
  \sigma_k^2(t)=\mean*{N_k^2}(t)-\mean*{N_k}^2(t)=\ton*{\frac{N^2}{M}-\frac{N^2}{M^2}}\ton*{1-\me^{-2\frac{t}{N}}}.
  \label{eq:variance_N_k}
\end{equation}
As also shown in Fig.~\ref{fig:large_c}, for small times and small
values of $\lambda$ our model with personalized information behaves as
the usual voter model.
This is due to the fact that for $t<\bar{t}$
the external information is completely random and so, being
all opinions equally numerous at the beginning, a voter model-like update
or a personalized information update are equivalent. This implies that we
can use Eqs.~\eqref{eq:drift_evolution_N_k} and
\eqref{eq:variance_N_k} to determine how $N_{\max}$ and thus $x_{\max}$
evolve for small times even in the presence of personalized
information
\[
N_{\max}(t)\approx\mean*{N_k(t)}+\sigma_k(t)\ \to\ x_{\max}(t)\approx\sigma_k(t).
\]
By expanding Eq.~\eqref{eq:variance_N_k} for small times we thus obtain
\[
x_{\max}(t)\approx\sqrt{2\frac{t}{N}\ton*{\frac{N^2}{M}-\frac{N^2}{M^2}}}\approx2\sqrt{\frac{N}{M}t} \ \text{  for  } t\leq\bar{t},
\]
where again we assumed $M\gg1$. Finally, we know from empirical
evidence that for large times it holds $x_{\max}(t)\sim t^{3/2}$; this
implies that the functional form of $x_{\max}(t)$ must be
\[
x_{\max}(t)=\alpha t^{3/2}+2\sqrt{\frac{N}{M}}t^{1/2}.
\]
The prefactor $\alpha$ can now be determined imposing the $t^{3/2}$
scaling to become dominant when the binomial scaling $t^{1/2}$ ends,
so for $t=\bar{t}=\sqrt{M}$. This gives
\[
\alpha \bar{t}^{3/2}=2\sqrt{\frac{N}{M}}\bar{t}^{1/2}
\]
and so
\[
\alpha = \frac{N^{1/2}}{M}.
\]
This shows that for sufficiently large $t$ it holds
\[
x_{\max}(t)\sim \frac{N^{1/2}}{M}t^{3/2}.
\]
\section{Evolution of $N_k-N/M$}
\label{apx:evolution_xk_yk}
The transition rate for $N_k$ in the presence of personalized information is
\[
W(N_k\to N_k+1)=(1-\lambda)\frac{N-N_k}{N}\frac{N_k}{N}+\lambda P_{PI,\to k},
\]
where the first term is the usual voter contribution,
while the second one is due to PI. Denoting as
$S_k=Ns_k$ the number of agents whose PI is polarized
along opinion $k$, we can write the latter as
\[
P_{PI\to k}=\frac{N-N_k-\sum_{j\neq k} S_k}{N}\frac{1}{M_s(t)}.
\]
Let us explain this expression. Among all the $N$ agents, those
already with opinion $k$ do not contribute to the transition rate to
opinion $k$, while those whose PI is polarized along $j\neq k$ can
make a transition toward $k$ only by a voter update. As a consequence
only $N-N_k-\sum_{j\neq k} S_k$ should be considered in computing the
transition probability. Moreover the PI of such agents will be
unpolarized and so we can assume that it suggests a random opinion,
giving the factor $1/M_s(t)$. Analogously
\[
W(N_k\to N_k-1)=(1-\lambda)\frac{N-N_k}{N}\frac{N_k}{N}+\lambda P_{PI, k\to}
\]
with 
\[
P_{PI, k\to}=\frac{N_k-S_k}{N}\frac{M_s(t)-1}{M_s(t)}.
\]
Inserting these transition rates in Eq.~\eqref{drift}
we can write the drift of $N_k$ as
\[
\nu_k(N_k) = \frac{\lambda}{M_s}\qua*{N-S-M_s\ton*{N_k-S_k}},
\]
where we introduced the total number of polarized agents as $S=\sum_kS_k$. 
The time evolution of $N_k$, neglecting diffusive fluctuations,
is thus
\[
\frac{dN_k}{dt}=\nu_k(N_k)=\lambda\qua*{\ton*{S_k-
    \frac{S}{M_s}}-\ton*{N_k-\frac{N}{M_s}}}.
\]
For $N\gg M$ and short times we can make the approximation $M_s\approx
M$ and so defining $y_k=S_k-S/M$ we arrive at an expression for the
time evolution of $x_k=N_k-N/M$
\begin{equation}
  \frac{dx_k}{dt}=\lambda\qua*{y_k-x_k}+\text{diffusive terms}.
  \label{eq:evolution_xk_yk}
\end{equation}
This expression provides additional support to our analytical
approach. Indeed, as shown in Fig.~\ref{fig:scaling_x_max}, it holds
$y_{max}(t)-x_{max}(t)\sim t^{1/2}$ and so Eq.~\eqref{eq:evolution_xk_yk}
predicts $x_{\max}(t)\sim t^{3/2}$, as actually observed in
Fig.~\ref{fig:scaling_x_max}. Note that by $y_{max}$ we denote $\max_k\qua*{y_k}$.

\section{Scaling regimes of the critical threshold}
\label{apx:scaling_regimes}

\subsection{$M\ll N$}
For $M\ll N$ during the first steps no opinion disappears and so we
can make the approximation $M_s(t)=M$. Inserting this into
Eq.~\eqref{eq:system} yields Eq.~\eqref{eq:system_M<<N}, that is
\begin{equation}
  \begin{dcases}
    t^*=\frac{4}{\lambda_c^2}\qua*{\sqrt{\frac{1-\lambda_c}{M}+\lambda_c}+\sqrt{\frac{1-\lambda_c}{M}}}^2\\
    \lambda_c = \frac{\gamma \frac{N^{1/2}}{M}(t^*)^{3/2}}{N+\gamma \frac{N^{1/2}}{M}(t^*)^{3/2}}.
  \end{dcases}
  \label{eq:system_M<<N_apx}
\end{equation}

From the expression for the polarization time $t^*$ we see that there
are two possible regimes.
\begin{itemize}
\item[\textbf{A)}] $\frac{1-\lambda_c}{M} \ll \lambda_c$\\
  The first equation in Eq.~\eqref{eq:system_M<<N_apx} becomes
  \[
  t^*=\frac{4}{\lambda_c}
  \]
  and replacing this expression into the second equation we obtain
  \[
  \lambda_c=\frac{\gamma N^{1/2}M^{-1}\ton*{\frac{4}{\lambda_c}}^{3/2}}{N+\gamma N^{1/2}M^{-1}\ton*{\frac{4}{\lambda_c}}^{3/2}}\approx 2N^{-1/2}M^{-1}\lambda_c^{-3/2}.
  \]
  This gives
  \[
  \lambda_c=2^{2/5}\ton*{NM^2}^{-1/5},
  \]
  so that
  \[
  \lambda_c M=2^{2/5}N^{-1/5}M^{3/5}.
  \]
  Our initial assumption $\lambda_c M\gg 1$
  is then verified provided that 
  \[
  M^{3/5}\gg N^{1/5}\to M\gg N^{1/3}.
  \]
  Recalling that we are assuming $M\ll N$ we then have
  \[
  \lambda_c=2^{2/5}\ton*{NM^2}^{-1/5} \ \text{for}\ N^{1/3}\ll M\ll N.
  \]
  Putting together the expressions for $t^*$ and $\lambda_c$ we finally obtain  Eq.~\eqref{eq:M<<N}
  \[
  \begin{cases}
    t^*=2^{8/5}\ton*{NM^2}^{1/5}\\
    \lambda_c=2^{2/5}\ton*{NM^2}^{-1/5}
  \end{cases}
  \]
  
\item[\textbf{B)}] $\frac{1-\lambda_c}{M}\gg\lambda_c$\\
  In this case the polarization time becomes
  \[
  t^*=\frac{16}{M\lambda_c^2}
  \]
  and again, by putting this expression into the expression for $\lambda_c$ we obtain
  \[
  \lambda_c=\frac{\gamma N^{1/2}M^{-1}2^6 M^{-3/2}\lambda_c^{-3}}{N+\gamma N^{1/2}M^{-1}2^6 M^{-3/2}\lambda_c^{-3}}\approx 2^4M^{-5/2}N^{-1/2}\lambda_c^{-3}.
  \]
  This gives
  \[
  \lambda_c=2\ton*{M^{5} N}^{-1/8}
  \]
  and so
  \[
  M\lambda_c\approx 2N^{-1/8}M^{3/8}.
  \]
  Consequently the two hypotheses $M\ll N$ and $M\lambda \ll 1$ are both satisfied if
  \[
  M \ll N^{1/3}
  \]
  and the expressions for $t^*$ and $\lambda_c$ are 
  \[
  \begin{cases}
    t^*\approx 4\ton*{MN}^{1/4}\\
    \lambda_c\approx 2\ton*{M^{5} N}^{-1/8}
  \end{cases}
  \]
  i.e., Eq.~\eqref{eq:M<<N^1/3}.
\end{itemize}

\subsection{$M=N$}
For $M=N$ one can no longer make the assumption $M_s(t)=M_s(0)=M$,
since some of the opinions disappear even during the first steps. In
this case we can approximated the decrease of $M_s(t)$ exploiting
Eq.~\eqref{eq:M_s_starnini}, which is exact for a simple multistate
voter model with $M=N$. The general expression is
\[
M_s(t) = \frac{M}{1+\frac{M}{N}t}
\]
and putting it into Eq.~\eqref{eq:system} we get
\begin{widetext}
  \begin{equation*}
    t^*=\frac{\pm\sqrt{\ton*{16\frac{\lambda N}{M}-8\lambda N - 16\frac{N}{M}}^2-64N\ton*{\lambda^2 N+16\lambda-16}}-\frac{16\lambda N}{M}+8\lambda N+16\frac{N}{M}}{2\ton*{\lambda^2N+16\lambda-16}},
    \label{eq:t_starnini_iniziale}
  \end{equation*}
  which, for $M=N$, becomes
  \begin{equation*}
    t^*=\frac{\pm\sqrt{\ton*{16\lambda-8\lambda N - 16}^2-64N\ton*{\lambda^2 N+16\lambda-16}}-16\lambda+8\lambda N+16}{2\ton*{\lambda^2N+16\lambda-16}},
    \label{eq:t_starnini_N=M}
  \end{equation*}
\end{widetext}
In the limit of large $N$ and small $\lambda$ this expression can be approximated as
\begin{equation}
  t^*=\frac{\pm32N^{1/2}+8\lambda N}{2\ton*{\lambda^2N-16}}
  \label{eq:t_starnini_approx}
\end{equation}
and, as it is possible to see, there are two distinct solutions. Requiring the denominator to vanish gives
\[
\bar{\lambda}^2N-16 = 0 \ \to \ \bar{\lambda} = \frac{4}{N^{1/2}}
\]
Note that while the solution with the plus, $t^*_{+}$, diverges in this limit, the one with the minus, $t^*_{-}$ is finite and positive, indeed 
\begin{align*}
  &\lim_{\lambda\to4N^{-1/2}}\frac{-32N^{1/2}+8\lambda N}{2\ton*{\lambda^2N-16}}=\\
  &=8N^{1/2}\lim_{\lambda\to4N^{-1/2}}\frac{\lambda N^{1/2}-4}{2\ton*{\lambda N^{1/2}+4}\ton*{\lambda N^{1/2}-4}}=\\
  &=\frac{N^{1/2}}{2}.
\end{align*}
Moreover, while for $\lambda<\bar{\lambda}$ $t^*_{-}$ is positive, the
other one is negative, meaning that $t^*_{+}$ is meaningful only in
the region $\lambda>\bar{\lambda}$ since a time must be a positive
quantity. For the solution with the minus we can then take the limit
$\lambda\to0$, which should give back the behavior of the multistate
voter model
\[
\lim_{\lambda\to0}t^*_{-}=\lim_{\lambda\to0}\frac{-32N^{1/2}+8\lambda N}{2\ton*{\lambda^2N-16}}=N^{1/2}.
\]
This result suggest that the solution $t^*_{-}$ is non physical, since
if $\lambda=0$ it holds $n_1=n_2$ (see Eqs.~\eqref{eq:mean_variance_1}
and \eqref{eq:mean_variance_2}) and so the two peaks should never
split meaning that $t^*=\infty$. In conclusion the expression for the
splitting time is the one with the plus and it is not defined for any
value of $\lambda$, more precisely
\begin{equation}
  t^*=\frac{32N^{1/2}+8\lambda N}{2\ton*{\lambda^2N-16}}\ \text{with}\ \lambda>\bar{\lambda}=\frac{4}{N^{1/2}}.
  \label{eq:t_starnini_finale}
\end{equation}
In the limit $\lambda\to 0$ and $N\to\infty$ we have two possible scaling regimes~\footnote{Only two because the scaling $\lambda<N^{-1/2}$ corresponds to negative times.}
\begin{itemize}
\item[\textbf{A)}] $\lambda N> N^{1/2} \ \to \ \lambda>N^{-1/2}$\\
  In this case we can approximate $t^*$ as 
  \[
  t^*\approx\frac{8\lambda N}{2\ton*{\lambda^2N-16}}\approx\frac{4}{\lambda}
  \]
  and substituting this expression into Eq.~\eqref{eq:condition_lambda_c_final} we get 
  \[
  \lambda_c\approx\ton*{\frac{4}{N^3}}^{1/5}\sim N^{-3/5},
  \]
  where we also exploited the fact that $M=N$.
  Note, however, that this result is in contrast with the initial assumption $\lambda>N^{-1/2}$ and so this scaling regime is impossible.
\item[\textbf{B)}] $\lambda\sim N^{-1/2}$ \\
  In this case we can set $\lambda^2N=16+\epsilon$ and Eq.~\eqref{eq:t_starnini_finale} becomes
  \[
  t^*\approx\frac{32N^{1/2}}{\epsilon}.
  \]
  Moreover, using Eq.~\eqref{eq:condition_lambda_c_final}, we obtain
  \[
  \lambda_c=\gamma N^{-3/2}\ton*{t^*}^{3/2}=2^{13/2}N^{-3/4}\epsilon^{-3/2}
  \]
  and setting $\lambda_c=\sqrt{\frac{16+\epsilon}{N}}$ we get
  \[
  \epsilon=2^{11/3}N^{-1/6}.
  \]
  This is consistent with $\epsilon$ being a small correction (that is our initial assumption) and moreover we see that it is the smaller the larger is $N$.
\end{itemize}
In conclusion we have
\begin{equation*}
  \begin{cases}
    t^*\approx2^{4/3}N^{2/3}\\
    \lambda_c\approx\frac{4}{N^{1/2}},
  \end{cases}
\end{equation*}
that is Eq.~\eqref{eq:M=N}.

\end{document}